# Ferroelectric switching of odd-parity magnon spin splitting

Yuhan Liang[1,2], Xingyu Yan[3], Bowen Hao[4,6], Ziye Zhu[4,6], Tianle Sui[3], Daniel Pharis[2], Xiaoxi Huang[2], Rakshit Jain[2,8], Tong Zhou[4,6†], Di Yi[3,†], Wanjun Jiang[5], Pu Yu[5], Igor Žutić[7], Yuan-Hua Lin[3], Daniel C. Ralph[2,8,†], Tianxiang Nan[1,†]

*1 School of Integrated Circuits and Beijing National Research Center for Information Science and Technology (BNRist), Tsinghua University, Beijing 100084, China*

*2 Department of Physics, Cornell University, Ithaca, NY 14853, USA*

*3 School of Materials Science and Engineering, Tsinghua University, Beijing 100084, China*

*4 Ningbo Institute of Digital Twin, Eastern Institute of Technology, Ningbo, Zhejiang 315200, China*

*5 Department of Physics, Tsinghua University, Beijing 100084, China*

*6 Department of Physics, University of Science and Technology of China, Hefei, Anhui 230026, China*

*7 Department of Physics, University at Buffalo, State University of New York, Buffalo, NY 14260, USA*

*8 Kavli Institute at Cornell for Nanoscale Science, Ithaca, NY 14853, USA*

[†] Email: tzhou@eitech.edu.cn; diyi@mail.tsinghua.edu.cn; dcr14@cornell.edu; nantianxiang@mail.tsinghua.edu.cn

**Magnetic symmetry can lift spin degeneracy in momentum space without producing net magnetization, generating spin textures classified by their parity under momentum reversal. Even-parity textures have well established in altermagnets. Odd-parity spin textures have recently emerged in electronic bands of compensated magnets, but their counterpart in collective bosonic excitations remains experimentally unresolved. Magnons provide a natural setting for this extension because they carry spin angular momentum through insulating magnets without accompanying charge flow. Beyond realizing this missing state, a broader challenge is to program spin splitting with voltage at room temperature. Here we report room-temperature transport evidence for ferroelectric switching odd-parity magnon spin splitting in multiferroic $BiFeO_3$. Symmetry analysis and spin-wave calculations reveal that the cycloidal chirality splits opposite-spin magnon branches with an odd-in-momentum dependence and sets the sign. Experimentally, an injected in-plane-polarized spin current generates an out-of-plane magnon spin component during propagation. This component exhibits the crystalline angular dependence predicted by theory and reverses upon ferroelectric switching, providing a transport fingerprint of the spin-split magnon state. We use the programmed magnon spin to drive deterministic field-free switching of a perpendicular ferromagnet and demonstrate XNOR logic-in-memory. Our results extend odd-parity spin splitting from fermionic electronic states to bosonic collective modes and establish nonvolatile electrical control at room temperature.**

Momentum parity provides a useful organizing principle for spin-split states in compensated magnets[1-3]. In even-parity states, the spin polarization is preserved between $+\mathbf{k}$ and $-\mathbf{k}$ ($\mathbf{k}$ is the momentum)[4-10], whereas in odd-parity states it reverses sign[11-15]. Even-parity electronic splitting is exemplified by altermagnets[16], while recent experiments have revealed odd-parity electronic spin textures in noncollinear magnets[14,15]. Such states establish an unconventional form of spin-momentum locking in electronic Bloch bands.

This progress raises a fundamental question: can odd-parity spin splitting extended from electronic quasiparticles to collective bosonic excitations? Magnons are such bosonic candidates that carry spin angular momentum through electrically insulating magnets without accompanying charge flow[17-19]. Magnetic-crystal and spin-space symmetries can lift the degeneracy between magnon modes carrying opposite spin angular momenta, generating momentum-dependent magnon spin splitting[20,21]. Experimental studies have begun to establish even-parity spin-split magnons in compensated magnets[22], but odd-parity magnon spin splitting remains experimentally unestablished. Its realization would extend spin-momentum locking to bosonic collective modes and introduce a new degree of control over spin transport in insulating magnets.

A further challenge is to make such a momentum-space spin structure electrically programmable. Multiferroics provide a natural route because their ferroelectric order can be coupled to the magnetic configuration responsible for spin splitting[23-30]. In the spin-spiral type-II multiferroic $NiI_2$, electric-field control of spiral chirality switches the associated odd-parity electronic spin polarization at cryogenic temperature[15]. Whether an analogous odd-parity magnon state can be realized, electrically reversed, and detected through spin transport at room temperature remains unknown.

The type-I multiferroic $BiFeO_3$ (BFO) provides a distinct platform for addressing these questions. BFO combines robust ferroelectricity with *G*-type antiferromagnetism at room temperature and hosts a long-wavelength cycloid spin texture whose propagation direction and handedness are coupled to the ferroelectric polarization[31]. Recent studies have identified momentum-dependent electronic spin textures in BFO, including *d*-wave[32] and *g*-wave[27,33] altermagnetic states in magnetic phases distinct from the long-wavelength cycloid studied here. Separately, experiments have established electric-field control of spin-wave frequencies[34,35], cycloidal order[36,37], magnon transport[29,38-40], and magnon torque[41,42]. However, these advances do not establish whether the ferroelectric polarization-locked cycloid can impose an odd-parity magnon splitting and whether this splitting can be reversed by switching ferroelectric polarization. Here we demonstrate room-temperature ferroelectric switching of odd-parity magnon spin splitting in BFO and use the electrically programmed magnon spin polarization to realize field-free magnetic switching and logic-in-memory functionality.

**Odd-parity magnon spin splitting in cycloidal BFO**

The momentum-space symmetry of a spin splitting is not restricted to electronic bands. For magnons, the analogous object is the signed frequency separation between collective modes carrying opposite spin angular momenta, $\Delta\omega(\delta\mathbf{k}) = \omega_{-}(\mathbf{k}_0 + \delta\mathbf{k}) - \omega_{+}(\mathbf{k}_0 + \delta\mathbf{k})$, where $\mathbf{k}_0$ is the high-symmetry nodal point in reciprocal space[43]. Magnetic symmetry determines both its momentum parity and its nodal structure. Even-parity splittings obey $\Delta\omega(-\delta\mathbf{k}) = \Delta\omega(\delta\mathbf{k})$; odd-parity splittings reverse sign. The simplest odd-parity bosonic state is *p*-wave[3], for which $\Delta\omega(\delta\mathbf{k}) = 2\chi v_p(\delta\mathbf{k}\cdot\mathbf{q}) + O(|\delta\mathbf{k}|^3)$, where $\chi$ is magnetic handedness and $v_p$ is splitting coefficient. Its defining signatures are therefore a linear splitting of opposite-angular-momentum modes, a nodal plane at $\delta\mathbf{k}\cdot\mathbf{q} = \mathbf{0}$, and reversal of the branch ordering under either momentum reversal or reversal of $\chi$. This provides a bosonic analogue of the *p*-wave spin-momentum locking recently established for electronic Bloch states, but now encoded in collective spin excitations.

Cycloidal BFO combines these ingredients with an electrically switchable order parameter. The *R3c* multiferroic BFO hosts *G*-type antiferromagnetic order and a robust ferroelectric polarization **P** ∥ <111>[44]. Magnetoelectric coupling generates a long-period cycloid (~ 62 nm) propagating along **q** ∥ <1$\bar{1}$0>, with spins rotating in the $\mathbf{P}$-$\mathbf{q}$ plane[36]. Its normal, $\widehat{\boldsymbol{D}} = (\mathbf{P}\times\mathbf{q})/|\mathbf{P}\times\mathbf{q}|$, sets the uniform cycloid-forming DM axis $\mathbf{D} = D_u\widehat{\boldsymbol{D}}$[45] and the dominant angular-momentum axis of the low-energy magnon modes. Consequently, reversing **P** at fixed **q** reverses the effective DM vector and cycloidal handedness[31] (Figs. 1a, 1b), providing a direct electrical handle on $\chi$. The magnetic symmetry of the cycloid then determines the magnon-band structure in momentum space. For a representative domain with $\mathbf{P}$ ∥ [111] and $\mathbf{q}$ ∥ [1$\bar{1}$0], cycloidal order breaks $\mathcal{PT}$ symmetry ($\mathcal{P}$ is space-inversion symmetry and $\mathcal{T}$ is time-inversion symmetry) while preserving the glide mirror $t_{1/2}\mathcal{M}_q$, composed of reflection across the plane $\mathcal{M}_q$ normal to $\mathbf{q}$ and a half translation $t_{1/2}$ along $\mathbf{P}$ (see Methods and Extended Data Figure 1). The glide mirror protects a degeneracy of the opposite-angular-momentum modes on $\delta\mathbf{k}\cdot\mathbf{q} = \mathbf{0}$, whereas away from this plane the reduced magnetic symmetry allows a handedness-dependent term linear in $\delta\mathbf{k}\cdot\mathbf{q}$. BFO therefore satisfies both defining requirements of a $p$-wave magnon state: a symmetry-protected nodal plane and a leading odd, linear-in-momentum splitting (Extended Data Table 1).

Our magnon spectrum calculations directly reproduce these features (see Methods). In the rotationally symmetric minimal model, the two branches are opposite circular precessions $\tau_D = \pm 1$ about $\mathbf{D}$, corresponding to opposite projections of mode spin angular momentum. They cross at the glide-plane-protected node and separate linearly with opposite ordering for $+\delta\mathbf{k}$ and $-\delta\mathbf{k}$ (Fig. 1c and Extended Data Figure 2). Switching the cycloidal handedness exchanges these angular-momentum-labelled branches but leaves the scalar, unpolarized spectrum unchanged. Thus, the electrically switchable variable is not the existence of the magnon bands themselves, but their momentum-dependent angular-momentum assignment. Because $\mathbf{D}$ is tilted out of the film plane, this $p$-wave magnon state additionally produces an out-of-plane spin-angular-momentum component, furnishing the transport signature measured below.

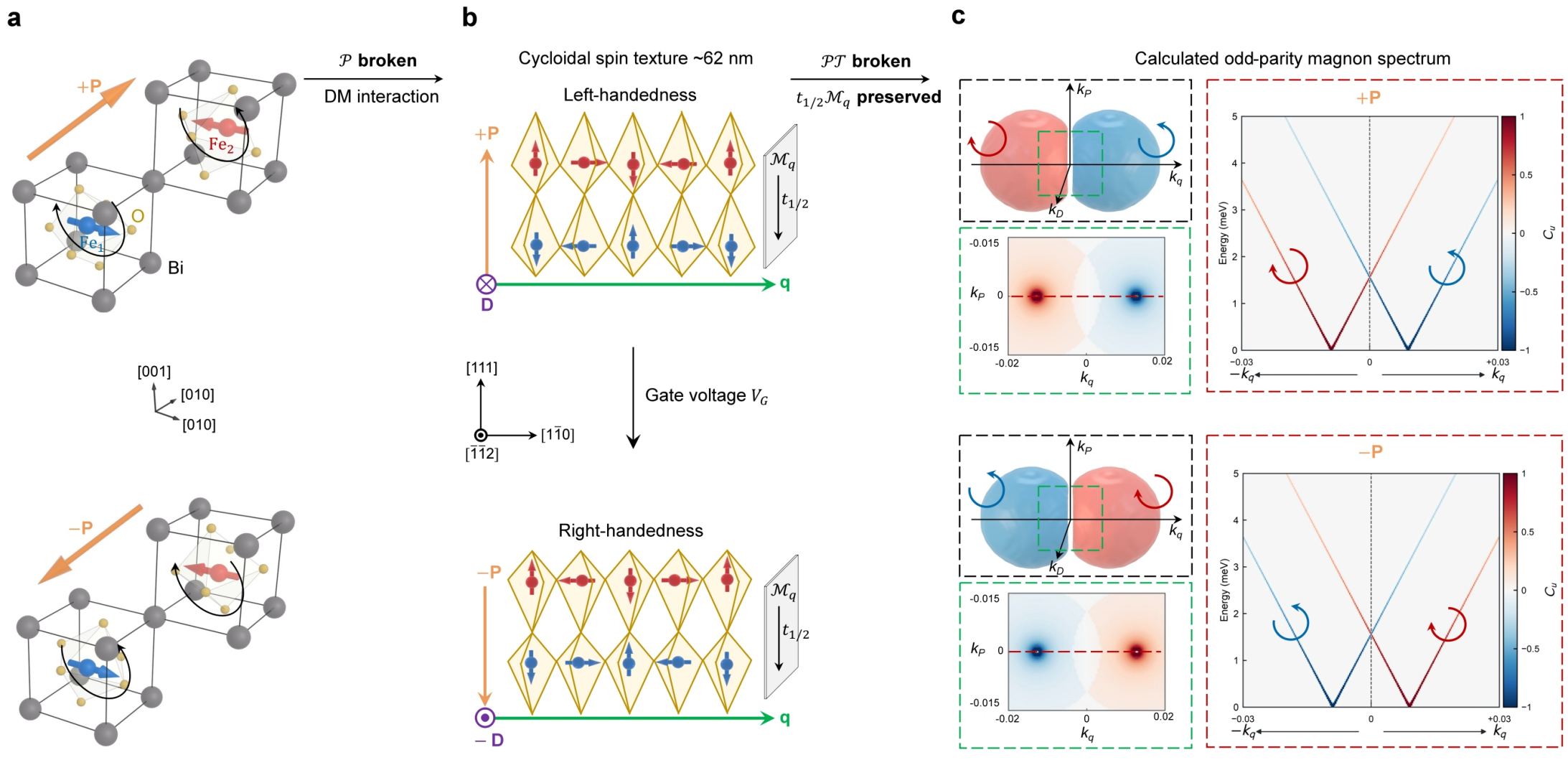

**Figure 1. Symmetry origin and calculated spectrum of odd-parity magnon spin splitting in BFO. a**, Schematic of the pseudo-cubic BFO unit cell, showing ferroelectric polarization $\mathbf{P}$ along one of the eight equivalent <111> directions. Blue and red spheres represent the two Fe sublattices ($Fe_1$ and $Fe_2$) with antiparallel magnetic moments, while gray and yellow spheres denote Bi and O atoms, respectively. An antiphase oxygen octahedra tilting (denoted by black arcs) is associated with $\mathbf{P}$, and is reversed upon switching $\mathbf{P}$. **b**, Schematic of cycloidal spin texture induced by the DM interaction due to the breaking of space-inversion symmetry $\mathcal{P}$. Upon $180°$ switching of $\mathbf{P}$, the effective DM vector $\mathbf{D}$ and cycloidal handedness are reversed. The antiphase oxygen-octahedral tilting breaks $\mathcal{PT}$ symmetry, and only the gliding mirror operation $t_{1/2}\mathcal{M}_q$ is preserved. **c**, Calculated odd-parity magnon response for the $+\mathbf{P}$ state (upper row) and the $-\mathbf{P}$ state (lower row). Here, $u$ denotes the unit vector normal to the cycloidal spin-rotation plane and parallel to $\mathbf{D}$. The black dashed boxes show the three-dimensional reciprocal-space response, the green dashed boxes show its two-dimensional $k_q$-$k_P$ cross-section, and the red dashed boxes show the corresponding energy–momentum cut along the cycloidal propagation direction $\mathbf{q}$. Red and blue indicate opposite signs of the calculated $C_u$ response, given in arbitrary units. The two low-energy branches carry opposite spin projections along $\mathbf{D}$. Reversing P reverses the $C_u$ pattern and exchanges the two circular-polarization channels, while leaving the scalar dispersion unchanged.

**Transport fingerprint of odd-parity magnon spin splitting**

Having established that cycloidal BFO can host odd-parity magnon spin splitting, we next ask how this band property can be detected. Directly resolving spin-split magnon branches by spectroscopy is challenging, especially in thin-film heterostructures and multidomain antiferromagnets[46]. We therefore inject a spin current with in-plane polarization $\boldsymbol{\sigma}_s$ and use the spin polarization $\boldsymbol{\sigma}_m$ carried by the resulting magnon current as a transport fingerprint. In a spin-degenerate antiferromagnetic channel, propagation may attenuate the injected spin, but $\boldsymbol{\sigma}_m$ remains parallel to $\boldsymbol{\sigma}_s$, without generating a transverse component (Fig. 2a). In spin-split BFO, by contrast, the magnon eigenmodes have a spin-quantization axis determined by the local $\mathbf{D}$ vector, which is tilted away from the film plane (*xy*-plane). The injected spin polarization $\boldsymbol{\sigma}_s$ therefore projects onto these spin-split eigenmodes, such that $\boldsymbol{\sigma}_m$ contains a component parallel to $(\boldsymbol{\sigma}_s \cdot \mathbf{D})\mathbf{D}$ as well as $\boldsymbol{\sigma}_s$. Consequently, the magnon current propagation along *z* acquires an out-of-plane component $\boldsymbol{\sigma}_m^z$, whose magnitude depends on $\boldsymbol{\sigma}_s \cdot \mathbf{D}$ (Fig. 2b and Methods). The emergence of this transverse spin-polarization component, together with its dependence on the relative orientation of $\boldsymbol{\sigma}_s$ and $\mathbf{D}$, thus provides a transport signature of magnon spin splitting.

To detect this magnon spin polarization, we fabricated 11 nm $SrRuO_3$ (SRO)/80 nm BFO/5 nm $Ni_{80}Fe_{20}$ (Py) heterostructures and performed spin-torque ferromagnetic resonance (ST-FMR) measurements (see Methods). The SRO layer serves as a high-symmetry spin-current source at room temperature[47,48]. An in-plane microwave charge current $J_c$ along $x$ generates a spin current flowing along the film normal $z$, with spin polarization $\boldsymbol{\sigma}_s \parallel y$. This spin current excites magnons in BFO, which propagate towards the Py detector and transfer their angular momentum to its magnetization. The component retaining the injected $y$-polarization produces a conventional damping-like torque[49], $\boldsymbol{\tau}_{DL}^{y} \propto \mathbf{M} \times (\mathbf{y} \times \mathbf{M})$ (where $\mathbf{M}$ is the magnetization orientation of the Py detector), with efficiency $\xi_{DL}^{y}$. Any out-of-plane magnon spin component generated during propagation would instead exert an unconventional torque, $\boldsymbol{\tau}_{DL}^{z} \propto \mathbf{M} \times (\mathbf{z} \times \mathbf{M})$, characterized by $\xi_{DL}^{z}$ (Fig. 2c). We patterned devices at crystalline angles $\varphi_E$, defined between $J_c$ and BFO [100]-direction (equivalently, between $\boldsymbol{\sigma}_s$ and [010]), to test the predicted angular dependence of the out-of-plane magnon polarization (Figs. 2d, 2e).

X-ray diffraction and reciprocal-space mapping confirm the epitaxial quality of both SRO and BFO layers (Extended Data Figures 3a, 3b). Piezoresponse force microscopy (PFM) identifies two ferroelectric domain variants and their polarization directions, while nitrogen-vacancy magnetometry verifies the corresponding cycloidal spin texture and propagation direction (Extended Data Figures 3c, 3d)[36]. All ST-FMR measurements were performed at room temperature, above the Curie temperature (~160 K) of SRO, thereby excluding contributions associated with its ferromagnetic order[50].

The ST-FMR response reveals a pronounced dependence of the out-of-plane magnon spin polarization on the angle of $\boldsymbol{\sigma}_s$ relative to the crystal axes (Fig. 2f). For devices with $\varphi_E = 0°$, i.e. $\boldsymbol{\sigma}_s \parallel [010]$, the ST-FMR spectra are consistent with a conventional torque $\boldsymbol{\tau}_{DL}^{y}$, where the symmetric $V_S$ and antisymmetric $V_A$ voltage Lorentzian components both simply reverse polarity upon magnetic-field reversal. When the device is rotated to $\varphi_E = 90°$, such that $\boldsymbol{\sigma}_s \parallel [\bar{1}00]$, the field-reversal symmetry of the Lorentzian components changes, indicating an additional torque $\boldsymbol{\tau}_{DL}^{z}$ arising from $\boldsymbol{\sigma}_m^{z}$. We separate the two torques by analyzing the magnetic-field angle $\varphi_B$ dependence of the ST-FMR[4] (see Methods and Extended Data Figure 4). Across devices with different $\varphi_E$, the extracted unconventional torque efficiency $\xi_{DL}^{z}$ follows a $sin\varphi_E$ dependence (Fig. 2g), whereas $\xi_{DL}^{y}$ is nearly isotropic, apart from a weak $cos^2\varphi_E$ modulation associated with the residual crystalline anisotropy of the SRO spin source (Extended Data Figure 5, Supplementary Note 1, Figs. S1 and S2 in Supplementary Materials)[42,47,48]. Measurements of 11 nm SRO/5 nm Py control sample further exclude SRO as the source of the unconventional spin torque (Extended Data Figure 6).

We interpret this crystalline-angle dependence of $\boldsymbol{\tau}_{DL}^{z}$ using the two-variant cycloidal variants in BFO. These variants have ferroelectric polarization $\mathbf{P}_1 \parallel [1\bar{1}\bar{1}]$ and $\mathbf{P}_2 \parallel [11\bar{1}]$, cycloid propagation direction $\mathbf{q}_1 \parallel [\bar{1}\bar{1}0]$ and $\mathbf{q}_2 \parallel [1\bar{1}0]$, and corresponding $\mathbf{D}$ vectors $\mathbf{D}_1 \parallel [\bar{1}1\bar{2}]$ and $\mathbf{D}_2 \parallel [\bar{1}\bar{1}\bar{2}]$. Because both cycloidal planes are tilted relative to the film plane, each variant can generate an out-of-plane magnon spin component from an injected in-plane polarization. For approximately equal populations of the two variants, the net out-of-plane polarization follows $\boldsymbol{\sigma}_m^z \propto \boldsymbol{\sigma}_s \cdot (\mathbf{D}_1 + \mathbf{D}_2)$. It vanishes for $\boldsymbol{\sigma}_s \parallel [010]$, for which $\boldsymbol{\sigma}_s \cdot (\mathbf{D}_1 + \mathbf{D}_2) = 0$, but become finite for $\boldsymbol{\sigma}_s \parallel [\bar{1}00]$, yielding the predicted $sin\varphi_E$ dependence (Fig. 2g; see Methods and Extended Data Figure 7). The agreement between this prediction and the measured $\boldsymbol{\tau}_{DL}^{z}$ provides the transport fingerprint of the odd-parity magnon spin splitting in cycloidal BFO.

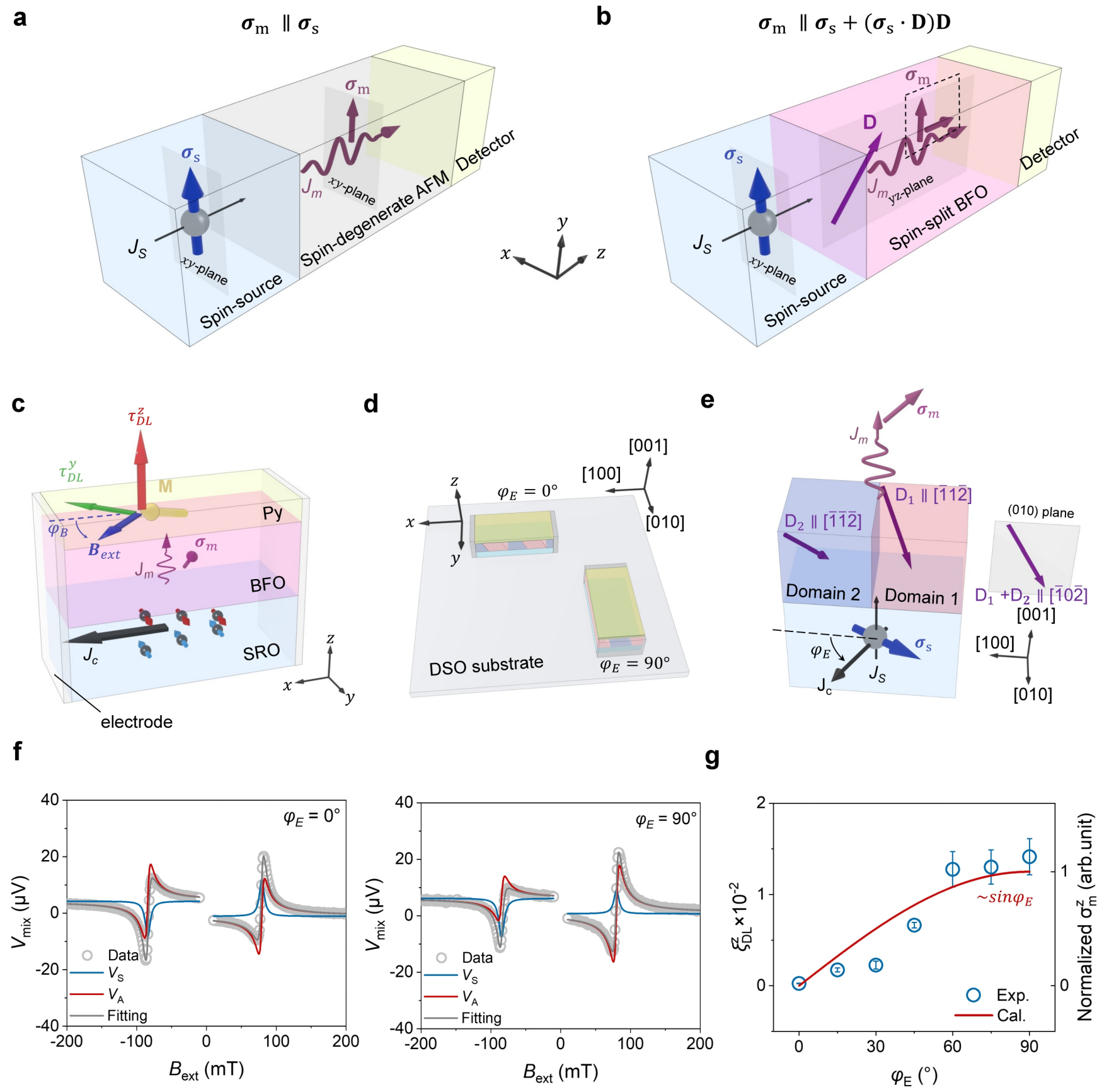

**Figure 2. Transport signature of magnon spin splitting in BFO. a**, Schematic of a *y*-polarized spin current $J_s$ injected into a spin-degenerate antiferromagnetic (AFM) magnon channel. The transmitted magnon spin polarization of magnon current $J_m$ remains collinear with the injected polarization $\boldsymbol{\sigma}_m \parallel \boldsymbol{\sigma}_s$. **b**, Schematic of a *y*-polarized spin current injected into a spin-split BFO magnon channel. The propagating magnon spin polarization tilts towards the local $\mathbf{D}$ vector and acquires an out-of-plane component $\boldsymbol{\sigma}_m^z$, with overall spin polarization $\boldsymbol{\sigma}_m$ containing components parallel to both $\boldsymbol{\sigma}_s$ and $(\boldsymbol{\sigma}_s \cdot \mathbf{D})\mathbf{D}$. The magnon spin polarization can be detected by a spin detector. **c**, ST-FMR measurement geometry. A microwave charge current $J_c$ flowing in the SRO layer generates a spin current along *z* with spin polarization along *y*, which is injected into BFO layer to excite a magnon current. The retained *y*-polarized magnon component generates a conventional torque $\boldsymbol{\tau}_{\mathrm{DL}}^y$ on the Py magnetization $\mathbf{M}$, whereas $\boldsymbol{\sigma}_m^z$ generates an unconventional torque $\boldsymbol{\tau}_{\mathrm{DL}}^z$. An external magnetic field $\mathbf{B}_{ext}$ is applied at an angle $\varphi_B$ to the *x*-axis. **d**, Schematic of two representative SRO/BFO/Py devices with $J_c$ applied at $\varphi_E = 0°$ and 90°. The BFO layer contains

two-variant domain structure. **e**, Schematic of interplay between spin current $J_s$ and DM vectors ($\mathbf{D}_1 \parallel [\bar{1}1\bar{2}]$ and $\mathbf{D}_2 \parallel [\bar{1}\bar{1}\bar{2}]$) in each domain for magnon current $J_m$ with magnon spin polarization $\boldsymbol{\sigma}_m$. The overall DM vector $\mathbf{D}_1 + \mathbf{D}_2 \parallel [\bar{1}0\bar{2}]$ lies in the (010) plane. The charge current $J_c$ is applied at an angle $\varphi_E$ with respect to the [100] direction. **f**, ST-FMR spectra for devices with $\varphi_E = 0°$ and $\varphi_E = 90°$, measured at $\varphi_B = 45°$. The experimental data are fitted to symmetric $V_S$ and antisymmetric $V_A$ voltage Lorentzians. **g**, Extracted $\xi_{DL}^z$ as a function of $\varphi_E$ (blue points). Error bars represent the standard error of torque efficiencies extracted for different frequencies. The theoretical calculation of normalized $\boldsymbol{\sigma}_m^z$ follows a $sin\varphi_E$ dependence (red line), showing good agreement with the experimental results.

**Ferroelectric switching of magnon spin polarization**

Having identified the out-of-plane magnon spin polarization $\boldsymbol{\sigma}_m^z$ as a transport fingerprint for odd-parity magnon spin splitting, we next test whether its sign can be reversed by ferroelectric switching. Because the SRO and Py layers in the ST-FMR geometry are electrically connected by shared electrodes, the BFO layer cannot be gated independently. We therefore fabricated *in-situ* gated 11 nm SRO/80 nm BFO/1.5 nm Ta/1 nm CoFeB/1.9 nm MgO devices with separate contacts to the SRO and the perpendicularly magnetized CoFeB detector. A $SiO_2$ layer electrically isolates the SRO drive channel from the CoFeB Hall circuit, enabling independent gating of BFO, current injection through SRO, and anomalous Hall readout from CoFeB (Fig. 3c; see Methods and Extended Data Figure 8). In this geometry, the unconventional torque $\boldsymbol{\tau}_{\mathrm{DL}}^z$ generated by $\boldsymbol{\sigma}_m^z$ is detected by a current-induced shift $\Delta\mathbf{B}_z$ of the CoFeB anomalous-Hall loop[51] (Fig. 3a). Reversing the ferroelectric polarization $\mathbf{P}$ reverses the cycloidal chirality [31], and is therefore expected to reverse the odd-parity magnon spin texture, $\boldsymbol{\sigma}_m^z$, and the corresponding sign of $\Delta\mathbf{B}_z$ (Fig. 3b).

Ferroelectric switching was verified through the ferroelectric diode effect in the device resistance[52] (see Methods and Extended Data Figure 9). Gate voltages of $V_G = \pm 2.8$ V produce two reproducible resistance states, corresponding to the $+\mathbf{P}$ and $-\mathbf{P}$ ferroelectric polarization states induced by preferred 180° ferroelectric switching[53]. The anomalous Hall loops measured for the two polarization states show nearly identical coercive fields and perpendicular magnetic anisotropy, excluding a voltage-controlled magnetic anisotropy effect [54] as the origin of the loop-shift reversal (see Methods and Extended Data Figure 10).

We then measured the anomalous Hall loops of the CoFeB under positive and negative bias currents $I_{\mathrm{bias}}$ applied to the SRO channel (see Methods). In the $-\mathbf{P}$ state set by $\mathrm{V_G} = -2.8$ V, opposite current polarities shifted the Hall loop in opposite directions, producing opposite signs of $\Delta\mathbf{B}_z$ consistent with the polarity of $\boldsymbol{\tau}_{\mathrm{DL}}^{z}$ inferred from the ST-FMR measurements (Fig. 3d). After switching the ferroelectric polarization to $+\mathbf{P}$ by applying $\mathrm{V_G} = +2.8$ V, the same current polarities produce reversed loop shifts, demonstrating reversal of $\Delta\mathbf{B}_z$ (Fig. 3e). The extracted $\Delta\mathbf{B}_z$ becomes detectable above a bias current of approximately 1 mA, consistent with the threshold current required for the out-of-plane torque to bias the magnetization reversal in perpendicular CoFeB[51]. At a fixed current polarity, $\Delta\mathbf{B}_z$ reverses deterministically upon ferroelectric switching (Fig. 3f), demonstrating the reversal of $\boldsymbol{\tau}_{\mathrm{DL}}^{z}$ and the underlying $\boldsymbol{\sigma}_m^z$. These results demonstrate *in-situ*, non-volatile ferroelectric control of the magnon spin associated with odd-parity magnon spin splitting, with direct electrical readout through anomalous-Hall loop shifts.

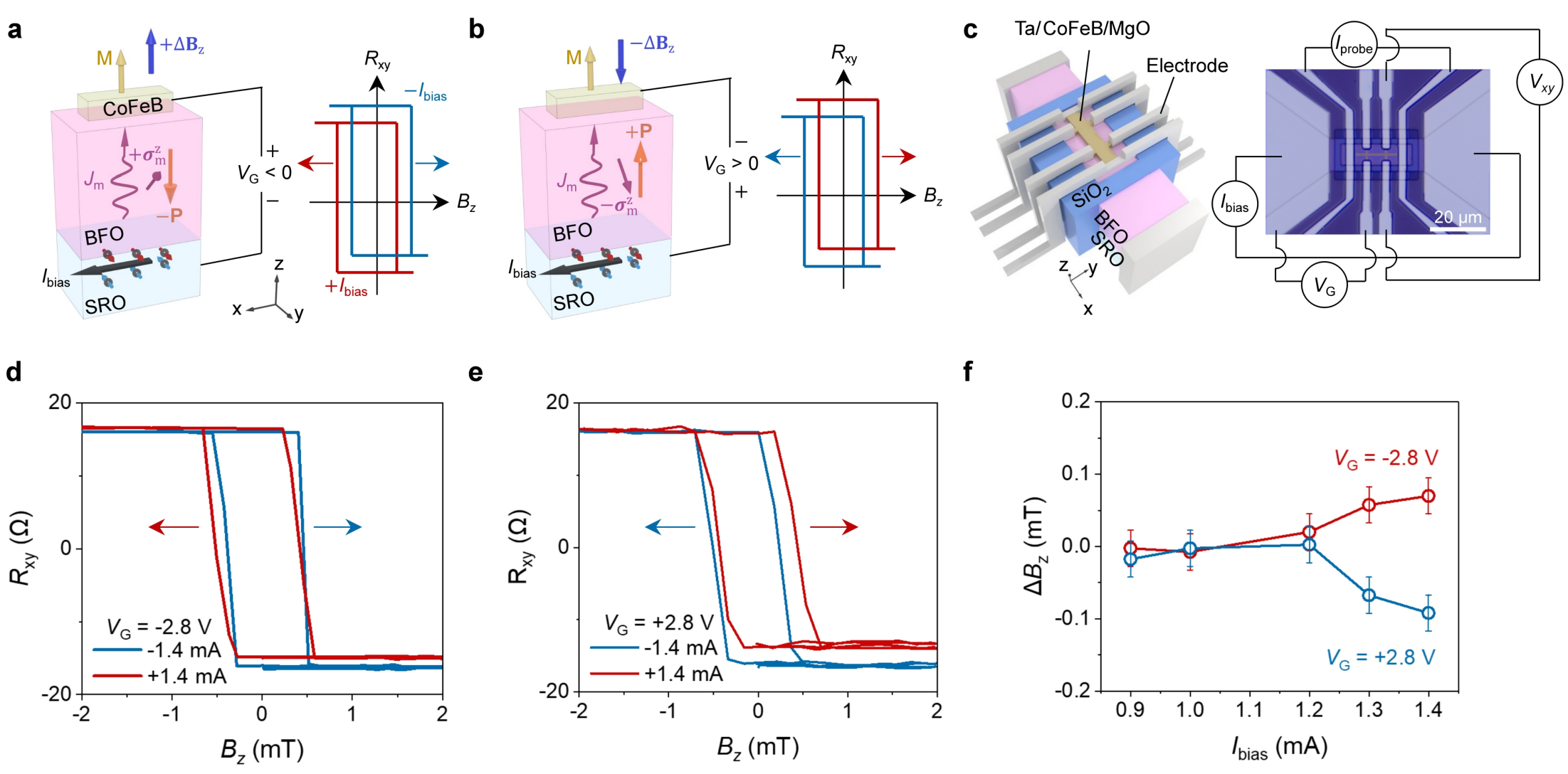

**Figure 3. *In-situ* electric-field control of out-of-plane magnon spin polarization in BFO. a** and **b**, Schematic of the *in-situ* gated measurement. **a**, A bias current $I_{\mathrm{bias}}$ flowing through the SRO layer injects a magnon current $J_{\mathrm{m}}$ into BFO with the $-\mathbf{P}$ state set by a negative gate voltage $\mathrm{V_G}$,

generating an out-of-plane magnon spin polarization $+\boldsymbol{\sigma}_m^z$. This exerts an out-of-plane torque on the top Ta/CoFeB/MgO structure, as read out by a current-induced shift $\Delta\mathbf{B}_z$ of the anomalous Hall loop. **b**, Applying a positive gate voltage $V_G$ switches the polarization to the $+\mathbf{P}$ state, which reverses the spin polarization component to $-\boldsymbol{\sigma}_m^z$ and correspondingly switches the sign of $\Delta\mathbf{B}_z$. **c**, Schematic of the in-situ device structure, false-colored optical image of the fabricated device, and electrical measurement configuration. A small probe current $I_{\text{probe}}$ = 10 μA is used to measure the anomalous Hall voltage $V_{xy}$. Scale bar, 20 μm. **d** and **e**, $R_{xy}$ loops of the top CoFeB layer measured under different $I_{\text{bias}}$ for (**d**) the $-\mathbf{P}$ state initialized by $V_G = -2.8$ V and (**e**) the $+\mathbf{P}$ state initialized by $V_G = +2.8$ V. Red and blue arrows indicate the loop-shift directions under $+I_{\text{bias}}$ and $-I_{\text{bias}}$, respectively. **f**, Extracted $\Delta\mathbf{B}_z$ as a function of $I_{\text{bias}}$ for the $-\mathbf{P}$ and $+\mathbf{P}$ states. Error bars are propagated from the coercive-field measurement resolution.

**Programmable field-free switching and logic-in-memory**

Finally, we use the ferroelectrically-switchable out-of-plane magnon spin $\boldsymbol{\sigma}_m^z$ and its associated unconventional torque $\boldsymbol{\tau}_{\text{DL}}^z$ to drive programmable field-free switching of the CoFeB in the same *in-situ* gated device (see Methods). Because $\boldsymbol{\tau}_{\text{DL}}^z$ provides an out-of-plane symmetry-breaking component, it enables deterministic switching of the perpendicular magnetization without an external symmetry-breaking magnetic field, a key requirement for high-density and energy-efficient magnetic devices[55]. The field-free switching polarity for a given sign of the pulse current $I_{\text{pulse}}$ is reversed between the $-\mathbf{P}$ and $+\mathbf{P}$ states of the BFO, due to the reversal of $\boldsymbol{\sigma}_m^z$ (Fig. 4a versus Fig. 4b).

Experimentally, in the $-\mathbf{P}$ state (set by $V_G = -2.8$ V), current pulses drive deterministic field-free switching of the perpendicular magnetization with a clockwise polarity and a critical switching current of approximately 2.8 mA, corresponding to a current density of $2.1\times10^6$ A/cm$^2$ (Fig. 4C). After switching BFO to the $+\mathbf{P}$ state (with $V_G = +2.8$ V), the same current-pulse sequence produces a counterclockwise loop (Fig. 4d). This reversal of the switching polarity is consistent with the ferroelectric reversal of $\boldsymbol{\sigma}_m^z$ and the corresponding $\boldsymbol{\tau}_{\text{DL}}^z$.

The ferroelectric state therefore nonvolatilely controls the magnetic switching polarity, enabling a logic-in-memory operation in a single device (see Methods). We use the gate voltage $V_G$ and the critical current-pulse $I_c$ polarity as two logic inputs and the final CoFeB magnetization, read through the anomalous Hall resistance $R_{xy}$, as the logic output. Sequential application of all four input combinations reproducibly generated the XNOR truth table (Figs. 4e, 4f). These results provide a proof-of-concept conversion of ferroelectrically programmable magnon spin splitting into field-free magnetic switching and logic-in-memory functionality.

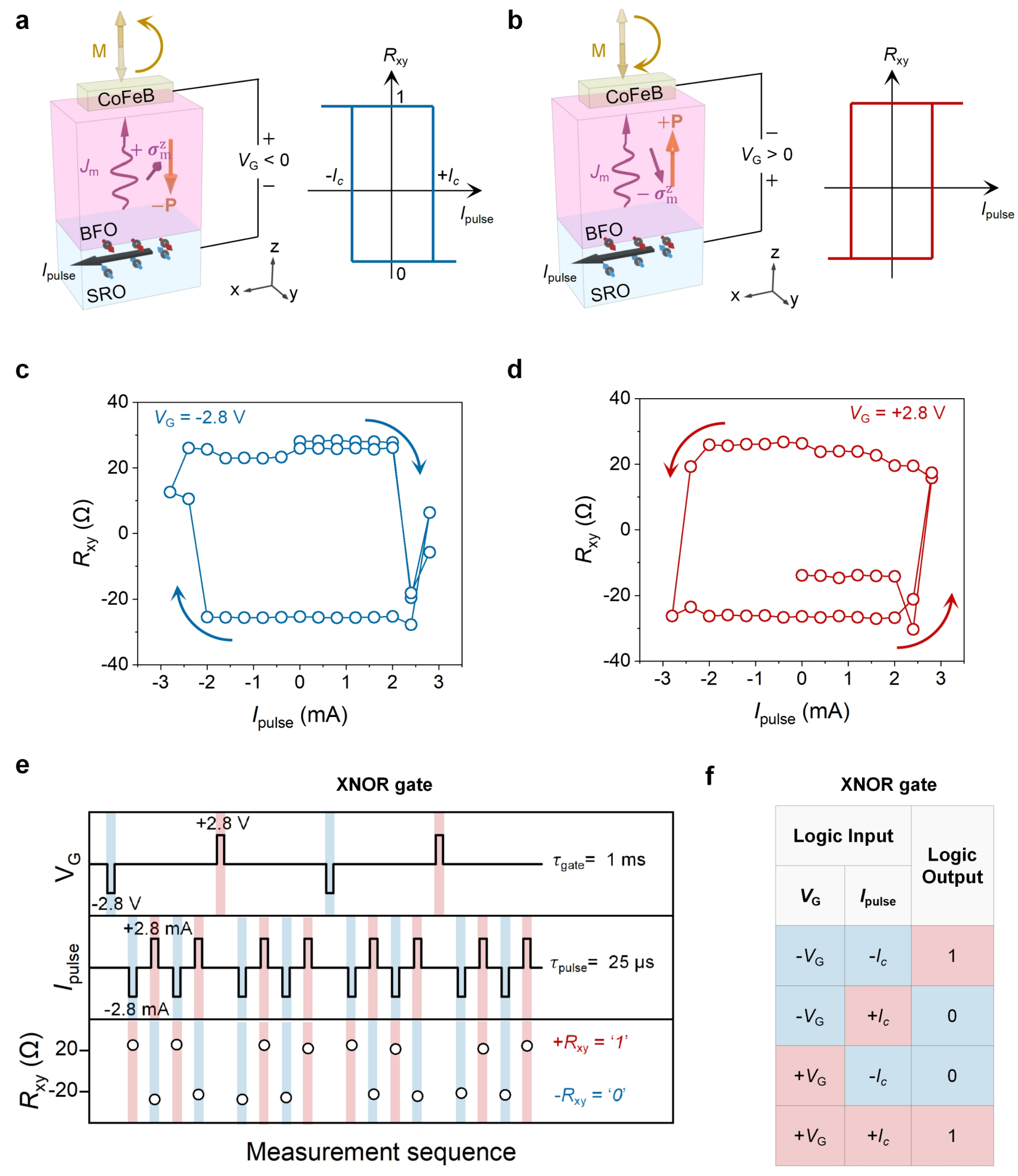

| Logic Input | | Logic Output |
|---|---|---|
| $V_G$ | $I_{pulse}$ | |
| $-V_G$ | $-I_c$ | 1 |
| $-V_G$ | $+I_c$ | 0 |
| $+V_G$ | $-I_c$ | 0 |
| $+V_G$ | $+I_c$ | 1 |

**Figure 4. Field-free switching and logic-in-memory driven by electrically programmable spin -torque. a** and **b**, Schematics of ferroelectrically programmed field-free magnetization switching polarity and logic-in-memory operation in an *in-situ* gated device. **a**, With BFO initialized in the **−P**

state, a pulse current $I_{\text{pulse}}$ applied to the bottom SRO generates a magnon current with $+\boldsymbol{\sigma}_m^z$ which switches the perpendicular CoFeB magnetization without an external magnetic field when $I_{\text{pulse}}$ exceeds critical current $I_c$. **b**, A positive gate voltage $\mathrm{V_G}$ switches BFO to the $+\mathbf{P}$ state, reversing the out-of-plane magnon spin to $-\boldsymbol{\sigma}_m^z$ and thereby reversing the magnetization switching polarity. The positive and negative anomalous Hall resistance $R_{\text{xy}}$ represent logic states 1 and 0, respectively. **c** and **d**, Current-induced field-free magnetization switching in the ferroelectric states set by $\mathrm{V_G} = -2.8$ V (**c**) and $+2.8$ V (**d**). **e**, Experimental XNOR logic-in-memory operation. The top two panels show the sequences of $\mathrm{V_G}$ ($\pm 2.8$ V) and $I_{\text{pulse}}$ ($\pm 2.8$ mA); the bottom panel shows the resulting $R_{xy}$, representing logic states 1 and 0. The gate- and current-pulse widths are 1 ms and 25 μs, respectively. **f**, Truth table of the XNOR logic gate controlled by $\mathrm{V_G}$ and $I_{\text{pulse}}$.

**Conclusion**

Our results establish ferroelectric switching of odd-parity magnon spin splitting as a route to electrically-programmable spin transport in insulating antiferromagnets. In BFO, switching the ferroelectric order reverses the cycloidal chirality and the sign of the associated magnon spin splitting. By identifying its crystalline transport fingerprint, we detect this state electrically via the out-of-plane spin-torque exerted on an adjacent ferromagnet and use the programmable torque to drive field-free magnetic switching and logic-in-memory operation. This work therefore extends spin-splitting functionality from electronic bands to electrically programmable odd-parity magnon states with spin-transport readout in a room-temperature multiferroic insulator (see Methods and Extended Data Table 2).

The present evidence is based primarily on spin-transport measurements; direct spectroscopic resolution of the switched magnon branches remains an important future direction[56]. Established approaches for controlling ferroelectric and antiferromagnetic domains[53], strain[57], thickness[58] and interface engineering[59] offer routes to strengthen and optimize the effect. Combined with theoretical predictions of altermagnetism in other BFO magnetic phases[27,29], these results point to a broader opportunity to connect electronic and magnonic spin splitting within a common multiferroic platform[60] and redefine the computing architecture to overcome the von Neumann bottleneck.

## Methods

### Magnetic symmetry analysis

Because magnetic symmetry operations act on the whole crystalline structure, when applying symmetry operations to the cycloidal spin texture we consider how the ferroelectric polarization **P**, cycloidal propagation direction $\mathbf{q}$, DM vector **D** and oxygen octahedra tilting transform simultaneously. In the paramagnetic phase, bulk BFO crystallizes in the rhombohedral *R3c* space group, hosting a three-fold rotation axis $\mathcal{C}_3 \parallel$ **P** and a triad of symmetrically equivalent $c$-glide planes generated by the $\mathcal{C}_3$ multiplication (here the *c*-axis is along **P**), among which is $t_{1/2}\mathcal{M}_q$ with a fractional lattice translation $t_{1/2} \parallel$ **P** and mirror operation $\mathcal{M}_q$ perpendicular to $\mathbf{q}$. After formation of a magnetic structure which has the cycloidal spin texture propagating along **q**, a spontaneous spatial symmetry breaking occurs, thereby $\mathcal{C}_3$ and the other two alternative glide planes are entirely extinguished, leaving $t_{1/2}\mathcal{M}_q$ the only allowed magnetic symmetry.

We then schematically show $t_{1/2}\mathcal{M}_q$ is a conserved operation (Extended Data Figure 1). The mirror operation $\mathcal{M}_q$ is applied first, which reverses **q**, **D** and the oxygen octahedra tilting along **P**. Importantly, $\mathcal{M}_q$ mirrors the atom positions along **q** and reverses the sublattice magnetization components parallel to $\mathcal{M}_q$ but conserves components perpendicular to $\mathcal{M}_q$. Here we conclude the chirality of cycloidal spin texture is conserved but the magnetic sublattices move by **q**/2 with reversal of the octahedra tilting. $t_{1/2}$ is then applied which moves the whole lattice along **P** by a half unit, which recovers the magneto-crystalline lattice. The net effect of $t_{1/2}\mathcal{M}_q$ is merely a reversal of the definitions of **q** and **D**, without changing the physical structure. We also examine the total DM interaction energy $\mathcal{H}_{DMI} = -\mathbf{D} \cdot (\mathbf{S}_i \times \mathbf{S}_j)$, where $\mathbf{S}_i$ and $\mathbf{S}_j$ are $Fe^{3+}$ spins along **q**. Under the $t_{1/2}\mathcal{M}_q$ operation, $\mathbf{D} \to -\mathbf{D}$ and $\mathbf{S}_i \times \mathbf{S}_j \to \mathbf{S}_j \times \mathbf{S}_i$, thereby $\mathcal{H}_{DMI}$ is invariant, consistent with the handedness conservation. We conclude that $t_{1/2}\mathcal{M}_q$ is a good symmetry for the cycloidal spin texture. These symmetry operations constrain the magnon spin quantization axis to lie along **D** and enforce an odd-parity magnon spin splitting, with a nodal plane perpendicular to **q** at $\mathbf{k} = 0$ (See Extended Data Table 1).

**Minimal microscopic model for odd-parity magnon polarization splitting**

The coordinate axes follow the experimental device geometry: $\boldsymbol{x} \parallel [100]$ and $\boldsymbol{y} \parallel [010]$ lie in the film, whereas $\boldsymbol{z} \parallel [001]$ is the film normal and the direction of vertical spin-current flow. We denote the ferroelectric polarization and cycloid wavevector by $\mathbf{P}$ and $\mathbf{q}$. In agreement with the main text, the magnon spin-quantization axis is:

$$\widehat{\boldsymbol{D}} = \frac{\boldsymbol{P} \times \boldsymbol{q}}{|\boldsymbol{P} \times \boldsymbol{q}|} \tag{1}$$

The rotating-frame code uses the antiparallel axis $\widehat{\boldsymbol{u}} = \boldsymbol{q} \times \boldsymbol{P}/|\boldsymbol{q} \times \boldsymbol{P}| = -\widehat{\boldsymbol{D}}$; this is only an axis convention. For a directed bond along $+\boldsymbol{q}$, the Dzyaloshinskii–Moriya (DM) vectors are:

$$\boldsymbol{D}_{ij} = D_u \widehat{\boldsymbol{D}}, \qquad \boldsymbol{D}_{ji} = -\boldsymbol{D}_{ij}. \tag{2}$$

Reversing the directed bond reverses its DM vector.

To isolate the long-period cycloid and its circular-polarization selection, we retain only nearest-neighbor and next-nearest-neighbor exchange and the uniform cycloid-forming DM interaction:

$$\mathcal{H} = J_1 \sum_{\langle ij \rangle} \boldsymbol{S}_i \cdot \boldsymbol{S}_j + J_2 \sum_{\langle \langle ij \rangle \rangle} \boldsymbol{S}_i \cdot \boldsymbol{S}_j + D_u \sum_{\langle ij \rangle_{\boldsymbol{q}}} \widehat{\boldsymbol{D}} \cdot (\boldsymbol{S}_i \times \boldsymbol{S}_j). \tag{3}$$

The alternating DM term that produces weak canting and the single-ion anisotropy are omitted. This is the smallest model with continuous rotation symmetry about $\widehat{\boldsymbol{u}}$ and therefore admits an exact

single-$\boldsymbol{k}$ rotating-frame spin-wave treatment. The approximation is used to determine symmetry, handedness and angular nodes, not an absolute transport coefficient.

The calculation uses an eight-Fe cell for the distorted pseudo-cubic structure, with lattice vectors (in Å):

$$\boldsymbol{a}_1 = (8.031984,0,0)$$
$$\boldsymbol{a}_2 = (0.119336,8.031098,0) \tag{4}$$
$$\boldsymbol{a}_3 = (0.119336,0.117576,8.030237)$$

The Fe basis contains 24 nearest-neighbor and 48 next-nearest-neighbor bonds. Their *G*-type antiferromagnetic parity and the antisymmetry of the DM vector under bond reversal were checked before the spin-wave calculation.

We use $S = 5/2$, $J_1 = 4.1\ meV$, $J_2 = 0.14\ meV$ and $|D_u| = 0.10\ meV$, corresponding to Model 2 of Ref.[61]. These are parameters of a minimal microscopic spin model rather than couplings extracted here by ab-initio total-energy mapping. To reproduce the frequency scale of the reference parameterization, the bilinear couplings in the spin-wave code are multiplied by:

$$\frac{\sqrt{S(S+1)}}{S} = 1.183215956620 \tag{5}$$

which leaves $D_u/J$, the period and all sign relations unchanged.

The generalized spiral is:

$$\boldsymbol{S}_{i,\boldsymbol{n}} = R_{\hat{\mathbf{u}}} \quad (2\pi\boldsymbol{k}\cdot\boldsymbol{n})\boldsymbol{S}_{i,0} \tag{6}$$

where $i$ labels the Fe basis and $\boldsymbol{n}$ a chemical-cell translation. Direct minimization gives:

$$\boldsymbol{k} = (0.0089784087, -0.0089784087,0)\ r.l.u., \qquad \lambda \simeq 628\ \text{Å}. \tag{7}$$

The reciprocal-lattice units here refer to the eight-Fe cell; equivalently, the pseudo-cubic wavevector is $\boldsymbol{q} = (\delta, -\delta, 0)$ with $\delta \simeq 0.004489$. The calculated period is consistent with the approximately 620 Å value[36]. Reversing $D_u$ reverses $\boldsymbol{k}$ without changing its magnitude.

Near the crossing of two opposite circular-polarization sectors, the minimal symmetry-allowed two-mode form is:

$$\mathcal{H}_{\text{eff}}(\delta\boldsymbol{k}) = \hbar\omega_0\mathbb{1} + \hbar v_2\delta k_\parallel^2\mathbb{1} + \chi\hbar v_p\delta k_\parallel\tau_D + O(\delta k^3), \qquad \delta k_\parallel = \delta\boldsymbol{k}\cdot\hat{\boldsymbol{q}} \tag{8}$$

Here $\tau_D = \pm 1$ labels opposite circular precessions about $\widehat{\boldsymbol{D}}$, and $\chi = \pm 1$ is the cycloidal handedness. Consequently,

$$\Delta\omega(\delta\boldsymbol{k}) = 2\chi v_p\delta k_\parallel = -\Delta\omega(-\delta\boldsymbol{k}) \tag{9}$$

which is the defining odd-in-momentum splitting. The effective model above is a low-energy symmetry expansion, while the coefficients and the full multi-band spectrum are obtained from the microscopic model.

Linear spin-wave theory for the single-$Q$ incommensurate spiral is evaluated in the generalized rotating frame following Ref.[62]. A momentum $\boldsymbol{Q}$ in the device coordinates receives contributions from the three sectors $\boldsymbol{Q}-\boldsymbol{k}$, $\boldsymbol{Q}$, and $\boldsymbol{Q}+\boldsymbol{k}$. The positive-frequency spin structure-factor tensor is:

$$S^{\alpha\beta}(\boldsymbol{Q},\omega) = \sum_n A_n^{\alpha}(\boldsymbol{Q})A_n^{\beta*}(\boldsymbol{Q})\delta[\omega-\omega_n(\boldsymbol{Q})]. \tag{10}$$

We choose orthonormal vectors $\boldsymbol{v}\parallel\boldsymbol{q}$ and $\boldsymbol{c}\parallel\boldsymbol{P}$ in the cycloid plane. The circular basis and signed chiral spectrum are:

$$\begin{aligned}\boldsymbol{e}_\pm &= (\boldsymbol{v}\pm i\boldsymbol{c})/\sqrt{2}, \quad I_\pm = \boldsymbol{e}_\pm^\dagger S\boldsymbol{e}_\pm,\\ C_u(\boldsymbol{Q},\omega) &= \mathrm{Im}[S_{vc}-S_{cv}] = I_- - I_+.\end{aligned} \tag{11}$$

$C_u$ is a signed circular-polarization diagnostic. It is not, by itself, a non-equilibrium magnon current. A thermodynamic mode magnetic moment would instead be defined from a field derivative:

$$s_{n,u} = -\frac{\partial(\hbar\omega_n)}{\partial B_u} \tag{12}$$

with modes followed through crossings by para-unitary eigenvector overlap. For quantitative claims, we do not replace the field-derivative expression above by an unnormalized structure-factor intensity.

To display the splitting directly, we follow the two isolated low-energy branches selected by $I_{u,+}$ and $I_{u,-}$ within $|t| \leq 0.006$ r.l.u. and $0.05 < \hbar\omega < 3$ meV. This restricted window avoids higher-band crossings at which a brightest-peak rule would change mode identity. Defining:

$$\Delta\omega(t) = \omega_{u,-}(t) - \omega_{u,+}(t) \tag{13}$$

we obtain:

$$\Delta\omega(t) =- \Delta\omega(-t), \qquad \Delta\omega(t; -D_u) =- \Delta\omega(t; +D_u). \tag{14}$$

The two branches cross at $t = 0$ and split linearly on either side, with $\partial\Delta\omega/\partial t =- 347.295$ meV/r.l.u. Both symmetry relations above hold to numerical precision: momentum reversal changes the sign of the splitting, whereas reversal of $D_u$ exchanges the two circularly polarized branches without changing their scalar dispersion (Fig. 1c and Extended Data Figure 2).

For completeness, the three theoretical views shown in the main-text Fig. 1 are generated from the same equilibrium linear-spin-wave calculation. The three-dimensional surface represents the signed circular-polarization response $C_u$ integrated over 0.05 < ħω < 2.0 meV; its red and blue lobes indicate opposite signs of $C_u$, and the isosurface level is a visualization threshold only. The two-dimensional map is a direct calculation in the orthogonal momentum coordinates $\mathbf{q} \parallel [1\bar{1}0]$ and $\mathbf{q} \perp [110]$, integrated over 0.05 < ħω < 5.0 meV. The energy-resolved panel samples $\boldsymbol{Q}(t) = [111] + \boldsymbol{t}[1\bar{1}0]$, where $t$ is the signed path coordinate in reciprocal-lattice units, and uses $C_u(\boldsymbol{Q}, \omega) = \mathrm{Im}[S_{vc}(t,\omega) - S_{cv}(t,\omega)] = I_-(t,\omega) - I_+(t,\omega)$. The opposite signs of the low-energy branches at the two propagation directions are the spectral manifestation of the momentum-odd chiral response. Reversing $D_u$ reverses $C_u$ and exchanges the two circular-polarization channels, while leaving the scalar dispersion unchanged. These panels are equilibrium spectral diagnostics and should not be interpreted as an absolute non-equilibrium magnon-transport coefficient.

**Theoretical calculation of unconventional magnon spin polarization induced by odd-parity magnon spin splitting**

The eigenmodes are quantized along $\widehat{\boldsymbol{D}}$, not along an independent $z$ axis. Because $\widehat{\boldsymbol{D}}$ is tilted out of the film, however, the two split branches carry opposite projections onto the device normal. For mode $n$ in domain $\ell$:

$$s_{n,z}(\boldsymbol{k}) = (\widehat{\boldsymbol{D}}_\ell \cdot \boldsymbol{z})\, s_{n,D}(\boldsymbol{k}) \tag{15}$$

Hence the same odd-in-$\boldsymbol{k}$ separation resolved in Extended Data Figure 2 is also a $z$-resolved spin splitting whenever $D_{\ell,z} \neq 0$; it is not a second splitting mechanism. For both experimentally observed domain variants below, $D_{\ell,z} =- 2/\sqrt{6}$, so each calculated pair of opposite-$D$ modes has finite and opposite $z$ polarization. The full calculation evaluates this projection from the Cartesian chiral vector $\boldsymbol{C} = (C_x, C_y, C_z)$, and Extended Data Figure 7 uses its calculated $C_z$ component in the two-domain sum.

The exact single-$Q$ rotating-frame implementation used here requires continuous spin-rotation symmetry about $\widehat{\boldsymbol{u}}$. Consequently, interactions that break this auxiliary continuous symmetry are not included in the numerical spectrum. This is a limitation of the computational representation, not an assumption required for the magnetic-symmetry classification.

The odd-parity conclusion follows from the discrete symmetries of the cycloidal state: they allow a leading term that is odd in $\delta k_{\parallel} = \delta \boldsymbol{k} \cdot \widehat{\boldsymbol{q}}$ and changes sign with the cycloidal handedness. Additional symmetry-preserving microscopic interactions may renormalize mode energies, velocities and polarization ellipticity, but do not remove this symmetry-allowed odd-in-momentum invariant. We therefore use the minimal $J_1$–$J_2$–$D_u$ model to expose the p-wave term and do not interpret it as a complete fit to every fine-structure feature of the BFO magnon spectrum.

The bottom $SrRuO_3$ layer generates an in-plane spin accumulation $\boldsymbol{\sigma}_s$. Its leading interfacial exchange coupling is:

$$\mathcal{H}_{int} =- g_{int}\, \boldsymbol{\sigma}_s \cdot \delta \boldsymbol{S} \tag{16}$$

so, injection depends on the overlap of the source spin with the magnon polarization. For one cycloidal domain the lowest-rank projector is $\widehat{\boldsymbol{D}}\widehat{\boldsymbol{D}}^{\mathsf{T}}$. This tensor fixes the angular nodes but is even under $\widehat{\boldsymbol{D}} \rightarrow - \widehat{\boldsymbol{D}}$. Electrical reversal therefore enters through the cycloidal handedness $\chi$:

$$J_\alpha^{\mathrm{odd}} = A_\alpha \chi\, D_\alpha(\boldsymbol{\sigma}_s \cdot \widehat{\boldsymbol{D}}) \tag{17}$$

The coefficients $A_\alpha$ collect interface transmission, propagation, relaxation, momentum filtering and detection. This response form is used only for the angular selection rule.

For an in-plane injected spin we define:

$$\boldsymbol{\sigma}_s(\varphi_E) = \sin\varphi_E\, \boldsymbol{x} + \cos\varphi_E\, \boldsymbol{y} \tag{18}$$

where $\varphi_E = 0$ corresponds to spin along $+[010]$ and positive angle rotates it toward $+[100]$. Any overall sign associated with the spin Hall source or voltage detector is included in $A_\alpha$.

The equilibrium calculation establishes the finite $z$ projection but does not determine the measured deflection angle: $A_z/A_y$ also depends on interface transmission, group velocity, lifetime and non-equilibrium occupation. The observed 14° angle is therefore not inferred from a bare geometric ratio.

**Theoretical calculation of unconventional magnon spin polarization induced by BFO with a two-variant domain structure**

The device contains two dominant variants:

$$\begin{aligned} \boldsymbol{P}_1 &\parallel [1\bar{1}\bar{1}], \quad \boldsymbol{q}_1 \parallel [\bar{1}\bar{1}0], \quad \widehat{\boldsymbol{D}}_1 = \frac{[-1,1,-2]}{\sqrt{6}}, \\ \boldsymbol{P}_2 &\parallel [11\bar{1}], \quad \boldsymbol{q}_2 \parallel [1\bar{1}0], \quad \widehat{\boldsymbol{D}}_2 = \frac{[-1,-1,-2]}{\sqrt{6}}. \end{aligned} \tag{19}$$

Thus $D_x$ and $D_z$ have the same sign in the two variants, while $D_y$ changes sign. For domain-1 weight $w$ and domain-2 weight $1-w$, the minimal handedness-odd response is:

$$J_\alpha^{\mathrm{odd}} = A_\alpha \chi \sum_{\ell=1}^{2} w_\ell\, D_{\ell,\alpha}(\boldsymbol{\sigma}_s \cdot \widehat{\boldsymbol{D}}_\ell) \tag{20}$$

For equal ideal domains:

$$\frac{1}{2}\sum_{\ell=1}^{2} \widehat{\boldsymbol{D}}_\ell \widehat{\boldsymbol{D}}_\ell^{\mathrm{T}} = \begin{pmatrix} 1/6 & 0 & 1/3 \\ 0 & 1/6 & 0 \\ 1/3 & 0 & 2/3 \end{pmatrix}, \qquad J_z^{\mathrm{odd}} = \frac{A_z \chi}{3} \sigma_x. \tag{21}$$

Therefore $[010] \to z$ cancels and $[100] \to z$ survives. Using the injected-spin definition above:

$$J_z^{\mathrm{odd}}(\varphi_E) = \frac{A_z\chi}{3}\sin\varphi_E. \tag{22}$$

This is the two-domain symmetry origin of the observed sinusoidal envelope. For unequal domain populations:

$$J_z^{\mathrm{odd}} = \frac{A_z\chi}{3}[\sigma_x + (1-2w)\sigma_y] \tag{23}$$

We next test this selection rule using the calculated low-energy spin-wave modes rather than the ideal unit-vector projector. For every domain $\ell$, mode $n$ and sampled momentum $\boldsymbol{Q}$, we calculate the chiral spectral vector in the device coordinates:

$$\boldsymbol{C}_{\ell n}(\boldsymbol{Q}) = \left(\mathrm{Im}[S_{yz} - S_{zy}], \mathrm{Im}[S_{zx} - S_{xz}], \mathrm{Im}[S_{xy} - S_{yx}]\right)_{\ell n \boldsymbol{Q}} \tag{24}$$

The low-energy polarization-response kernel is constructed as:

$$R_{\alpha\beta}^{\mathrm{SW}}(w) = \sum_{\ell,n,\boldsymbol{Q}}^{\hbar\omega<5\ meV} w_\ell \ \frac{C_{\ell n,\alpha}(\boldsymbol{Q})C_{\ell n,\beta}(\boldsymbol{Q})}{|\boldsymbol{C}_{\ell n}(\boldsymbol{Q})|} \tag{25}$$

Each mode therefore contributes the dyadic $\boldsymbol{C}_{\ell n}\boldsymbol{C}_{\ell n}^{\mathrm{T}}/|\boldsymbol{C}_{\ell n}|$. The plotted quantity is:

$$J_z^{\mathrm{SW}}(\varphi_E, w) = [R^{\mathrm{SW}}(w)\boldsymbol{\sigma}_s(\varphi_E)]_z \tag{26}$$

normalized by $|R_{zx}^{\mathrm{SW}}(w=1/2)|$. Because the dyadic kernel is even under $\boldsymbol{C} \to -\boldsymbol{C}$, it determines the angular envelope and relative domain cancellation but not the switchable overall sign. The latter is the handedness-odd factor $\chi$ already displayed explicitly in the two-domain response above; Extended Data Figure 7 is plotted for a fixed sign of $\chi$.

Using the distorted BFO cell rather than ideal pseudo-cubic unit vectors gives:

$$R_{\mathrm{geo}} = \begin{pmatrix} 0.168273 & -0.000012 & 0.334119 \\ -0.000012 & 0.168273 & -0.002488 \\ 0.334119 & -0.002488 & 0.663454 \end{pmatrix}, \qquad \left|\frac{R_{zy}}{R_{zx}}\right| = 7.45 \times 10^{-3}. \tag{27}$$

The small residual $[010] \to z$ term is therefore a lattice-metric correction, not the principal response. The calculated spin-wave kernel gives:

$$\left|\frac{R_{zy}^{\mathrm{SW}}}{R_{zx}^{\mathrm{SW}}}\right| = 7.4474 \times 10^{-3}, \qquad \left\| \frac{R^{\mathrm{SW}}}{\| R^{\mathrm{SW}} \|} - \frac{R_{\mathrm{geo}}}{\| R_{\mathrm{geo}} \|} \right\| = 6.1 \times 10^{-8}. \tag{28}$$

Thus the domain-resolved low-energy spin-wave calculation independently reproduces the angular node and relative channel cancellation (Extended Data Figure 7). Because the kernel contains no group velocity, lifetime, Bose weighting or interface transmission, it is an equilibrium polarization-response proxy rather than an absolute transport coefficient.

**Sample growth**

Epitaxial $SrRuO_3$ (SRO)/$BiFeO_3$ (BFO) heterostructures were deposited on orthorhombic (o) $DyScO_3$ (DSO) substrates with $(110)_o$-orientation, using pulsed laser deposition (PLD) with a 248 nm KrF excimer laser. SRO and BFO were deposited at substrate temperature of 670 °C and 700 °C, under an oxygen partial pressure of 110 mTorr and 150 mTorr, respectively. The laser fluence at the target surfaces was approximately 1.5 J/cm$^2$ with a pulse repetition rate of 5 Hz. Subsequently, the samples were cooled to room temperature in an oxygen rich atmosphere and transferred through air to a magnetron sputtering chamber with a base pressure of $2\times10^{-8}$ Torr for the deposition of ferromagnetic metals. The ferromagnetic $Ni_{80}Fe_{20}$ (Py) layer was deposited at an argon pressure of 3 mTorr, followed by the deposition of a 3 nm Ti capping layer to prevent oxidation. The 11 nm SRO/80 nm BFO/5 nm Py/3nm Ti heterostructure was employed for spin-torque ferromagnetic resonance (ST-FMR) measurements. We observed that the out-of-plane magnon spin polarization is weak for heterostructures with thinner BFO[42]. The 1.5 nm Ta/1 nm CoFeB/1.9 nm MgO/3 nm Ta multilayers with perpendicular magnetic anisotropy (PMA) were deposited on 11 nm SRO/80 nm BFO for loop shift and magnon-torque-switching measurements. Film thicknesses were measured using X-ray reflectivity. The high quality of epitaxy for the BFO thin films was characterized by X-ray diffraction (Extended Data Figure 3a) and reciprocal space mapping (Extended Data Figure 3b). The ferroelectric domain structure was characterized by piezoresponse force microscopy (PFM) (Extended Data Figure 3c).

**Nitrogen-vacancy (NV) center measurements**

Scanning Nitrogen-vacancy magnetometry (SNVM) was performed at room temperature using a commercial scanning NV microscope (QDAFM, CIQTEK). A diamond tip hosting a single negatively charged $NV^-$ center (a defect consisting of substitutional nitrogen atom adjacent to a carbon vacancy) was used to probe the weak stray magnetic field of BFO. The electronic ground state of the NV center is a spin triplet ($m_s = 0, \pm 1$). The NV center was continuously excited with a ~520 nm laser. A microwave source was applied to drive electron spin resonance (ESR) transitions between the spin sublevels ($m_s = 0 \rightarrow \pm 1$). Since the $m_s = \pm 1$ sublevels exhibit a lower probability of emitting photons during the deexcitation process, a resonant microwave causes a detectable reduction in the photoluminescence (PL) intensity. To resolve the two distinct ESR transitions, a small magnetic field with a projection along the NV axis was applied to lift the degeneracy of the $m_s = \pm 1$ sublevels, resulting in two characteristic dips in the PL-detected ESR spectrum. The spatial magnetic images were then acquired in the "iso-B" mode, which fixes the microwave frequency on the flank of one ESR dip (typically near its full width at half maxima). As the tip scans across the sample surface, the local stray magnetic field from BFO shifts the ESR frequency via the Zeeman effect, thereby modulating the PL intensity and generating magnetic contrast. The results are shown in Extended Data Figure 3b.

**Device fabrication**

The devices for ST-FMR measurements were 16 μm-wide and 80 μm-long microstrips with ground-signal-ground (GSG) electrodes using standard photolithography, followed by argon ion milling. Subsequently, 5 nm Ti / 200 nm Pt bilayers were deposited for electrode contacts using standard lift-off process. Devices for *in-situ* electric-field-controlled magnon spin splitting were fabricated in four steps: (1) The heterostructures were patterned into 12 μm-wide and 34 μm-long microstrips using standard photolithography, followed by argon ion milling. (2) A small Hall bar about 1 μm-wide and 20 μm-long, was patterned on top of the microstrips and carefully ion-milled to stop precisely at the BFO surface. The Hall detect lead is about 1 um-wide. (3) A 150 nm insulating $SiO_2$ layer was deposited via e-beam evaporation using a standard lift-off process. (4) Finally, 5 nm Ti /

400 nm Pt bilayers were deposited for electrode contacts using a standard lift-off process. See Extended Data Figure 8.

**ST-FMR measurements**

For ST-FMR measurements, a Keysight E8257D microwave generator was used to provide a 15 dBm microwave excitation, modulated by an internal oscillator at 1713 Hz. The mixing voltage signal was detected using a DSP 7265 lock-in amplifier. Magnetic fields were applied using a GMW projected-field magnet mounted on a Newport motion-controlled stage. The alternating magnon torque acting on the Py layer induces magnetization procession and leads to resistance oscillation due to the anisotropic magnetoresistance of the device. Mixing between the device resistance oscillation and $I_{RF}$ generates a resonance in the d.c. voltage $V_{mix}$, expressed as[63]:

$$V_{mix} = V_S \frac{\Delta^2}{(B - B_0)^2 + \Delta^2} + V_A \frac{\Delta(B - B_0)}{(B - B_0)^2 + \Delta^2} \tag{29}$$

where $\Delta$ is the resonant linewidth, and $B_0$ is the magnet resonant field. $V_S$ and $V_A$ are the symmetric and antisymmetric coefficient of the Lorentzian functions, related to the in-plane and out-of-plane magnon torques, respectively.

If the magnon carries spin polarization components $\sigma_x$, $\sigma_y$ and $\sigma_z$, along the *x*, *y* and *z* axis, respectively, the induced magnon torque has components along all of the *x*, *y* and *z* axis. The corresponding angle $\varphi_B$ dependences for the $V_S$ and $V_A$ coefficients are expressed as [4]:

$$V_A = V_{FL}^x sin\varphi_B sin2\varphi_B + V_{FL}^y cos\varphi_B sin2\varphi_B + V_{DL}^z sin2\varphi_B \tag{30}$$

$$V_S = V_{DL}^x sin\varphi_B sin2\varphi_B + V_{DL}^y cos\varphi_B sin2\varphi_B + V_{FL}^z sin2\varphi_B \tag{31}$$

Here, $V_{DL}^x$, $V_{DL}^y$ and $V_{DL}^z$ are coefficients for the damping-like torque generated by the $\sigma_x$, $\sigma_y$ and $\sigma_z$, respectively. $V_{FL}^x$, $V_{FL}^y$ and $V_{FL}^z$ are the field-like-torque counterparts. We note that $V_{FL}^y$ in principle can have contributions from both a field-like magnon torque and a current-induced Oersted field. An angle offset is also introduced to compensate magnetic field misalignment.

**Ferroelectric diode effect measurements**

The ferroelectric diode effect was employed to confirm ferroelectric switching in the fabricated *in-situ* electric-field controlled device. A series of gate voltage pulses ($V_G$) from +2.8 V to -2.8 V are applied with a pulse width of 1 ms using a Keithley 2400 source meter (schematic in Extended Data Figure 9a). The positive direction of voltage is defined from bottom electrode (SRO layer) to top electrode (CoFeB layer) (schematic in Extended Data Figure 9b). Following the removal of $V_G$ and a 4 s stabilization delay, a small readout voltage $V_{\text{probe}}$ = 0.5 V is applied to monitor the remanent leakage current ($I_{\text{leak}}$) using a Keithley 2400 source meter. We observed a hysteresis loop of $I_{\text{leak}}$ (Extended Data Figure 9c) and conclude that the ferroelectric polarization is switched by applying $V_G = \pm 2.8$ V.

***In-situ* electric-field controlled loop shift measurements**

For *in-situ* electric-field-controlled loop-shift measurements, a voltage pulse of $\mathrm{V_G} = \pm 2.8$ V with a 1 ms width was first applied to switch the ferroelectric polarization. Subsequently, a small probe current of $I_{\text{probe}}$ = 10 μA was applied by a Keithley 2400 source meter, while the Hall voltage was monitored using a Keithley 2182A nanovoltmeter. Concurrently, a bias current $I_{\text{bias}}$ was applied via another Keithley 2400 to generate the magnon current. These measurements were conducted within a Physical Property Measurement System (PPMS) under out-of-plane magnetic field sweeping. The value of $\Delta\mathbf{B}_z$ can be estimated via:

$$\Delta\mathbf{B}_z = -\frac{(B_{c+}^{+bias} + B_{c-}^{+bias}) - (B_{c+}^{-bias} + B_{c-}^{-bias})}{4} \tag{32}$$

Where $B_{c+}^{+bias}$ and $B_{c-}^{+bias}$ is the positive and negative coercive field under positive bias current, $B_{c+}^{-bias}$ and $B_{c-}^{-bias}$ is the positive and negative coercive field under negative bias current, respectively. The error bar of $\Delta\mathbf{B}_z$ is calculated as $B_{reso}$ / 2 = 0.025 mT, where $B_{reso}$ = 0.05 mT is the measurement resolution.

**Absence of voltage-controlled magnetic anisotropy (VCMA) effect**

To exclude the VCMA effect, we monitored the anomalous Hall resistance loop after applying a gate voltage $V_G = \pm 2.8$ V to switch the ferroelectric polarization. We found that the anomalous

Hall loops measured after the two gate polarities show nearly identical coercive fields and perpendicular magnetic anisotropy (Extended Data Figure 10), excluding the VCMA effect as the origin of the loop-shift reversal.

***In-situ* electric-field controlled field-free magnetization switching**

For both the *in-situ* electric-field control of field-free magnon-torque-induced magnetization switching and the logic-in-memory measurements, a polarization-switching gate voltage pulse ($V_G$ = ±2.8 V, 1 ms width) was applied prior to the current-pulse measurements. A Keithley 6221 source meter was then used to generate a single square-wave pulse with a 50% duty cycle. For the magnetization-switching measurements, the pulse period was 100 μs, corresponding to a 50 μs pulse duration, whereas for the logic-in-memory measurements, the pulse period was 50 μs, corresponding to a 25 μs pulse duration. After each current pulse, the Hall voltage was measured following an 8 s delay using a Keithley 2182A nanovoltmeter under a probe current $I_{\text{probe}}$ = 10 μA.

**Benchmark of odd-parity magnon spin splitting controlled by ferroelectric polarization in BFO**

Recent discoveries of unconventional compensated magnets have demonstrated that appropriate magnetic and crystalline symmetries can lift spin degeneracy without introducing macroscopic magnetization, thereby expanding the landscape of spintronic materials[2]. A prominent example is altermagnetism, where the alternating rotation of magnetic sublattices in the crystal structure breaks the spin degeneracy and gives rise to momentum-dependent spin splitting. Prototype altermagnets, such as $RuO_2$[4,6], exhibit characteristic even-parity spin splitting, which has subsequently been identified in a broad range of collinear magnetic materials, including CrSb[64], FeS[65], and MnTe[5,22,66], through both electronic and magnonic band structures.

However, the symmetry of collinear magnetic systems fundamentally restricts the spin splitting to even parity, motivating the exploration of odd-parity spin splitting in compensated magnets with noncollinear magnetic order. This possibility has attracted renewed attention following the emergence of altermagnetism and the realization that the interplay between crystalline symmetry and complex magnetic order can generate unconventional momentum-dependent spin splitting. Earlier studies of noncollinear Kagome antiferromagnets, such as $Mn_3X$ (X=Pt[67], Ir[68], Sn[69], etc.), revealed spin-polarized electronic states arising purely from their chiral noncollinear magnetic structures. This question has been revisited in light of the recent development of altermagnetism, where symmetry analyses of the coupled crystalline and magnetic structures have revealed new possibilities for odd-parity spin splitting. Experimentally, $MnTe_2$[10] was found to exhibit an even-parity plaid-like spin splitting, while odd-parity electronic spin splitting has been reported in $Gd_3(Ru_{1-\delta}Rh_\delta)_4Al_{12}$[14] and $NiI_2$[15] with commensurate or incommensurate spin spirals. Nevertheless, the magnonic counterpart of odd-parity spin splitting remains unexplored.

A remaining challenge beyond the discovery of new spin-split states is to achieve their voltage control, readout, and functional utilization for spintronic applications[26]. Multiferroic materials have been theoretically proposed as a promising platform for electrically manipulating spin splitting[24,25], yet experimental realization remains limited. Recent progress includes ferroelectric switching of electronic spin splitting in the type-II multiferroic $NiI_2$ at cryogenic temperatures[15]. However, an electrically controllable magnonic spin splitting, particularly with odd parity, has remained unexplored. Here, by combining symmetry analysis, spin-wave calculations, and room-temperature transport measurements, we demonstrate ferroelectric control of odd-parity magnon spin splitting in the canonical type-I multiferroic BFO (Figs. 1–4 in the main text). The key scientific advances of this work are summarized in Extended Data Table 2.

## Acknowledgments

**Funding:** T.N. acknowledges National Key R&D Program of China under Grant number 2024YFB3614100. D.Y. acknowledges National Natural Science Foundation of China under Grant number 52650104. T.Z. acknowledges the Zhejiang Provincial Natural Science Foundation of China (LR25A040001), the Zhejiang Provincial Leading Innovative and Entrepreneurial Team Project (2025R01017), the National Natural Science Foundation of China (12474155, 12688201), and the Zhejiang Leading Goose Project (2026C02A2013(SD2)). I. Ž. acknowledges US Department of Energy (DOE), Office of Science, Office of Basic Energy Sciences (BES), under award number DE-SC0004890. D.C.R. gratefully acknowledges US Department of Energy (DOE), Office of Science, Office of Basic Energy Sciences (BES), under award number DE-SC0017671 for measurements at Cornell. The research made use of the shared facilities of the Cornell NanoScale Facility, a member of the National Nanotechnology Coordinated Infrastructure (supported by the US National Science Foundation via grant NNCI-2025233) and the facilities of the Cornell Center for Materials Research.

**Author contributions:** T.N. and D.C.R. conceived and supervised the experiments. Y.L., X.Y., T.S., D.P. performed the sample growth. Y.L. performed the device fabrication. Y.L., X.H. and R.J. performed device measurements and analysis. B.H., Z.Z., and T.Z. conducted theoretical calculations. T.N., D.C.R., D.Y., T.Z., Y.L., and B.H. wrote the manuscript. Y.-H.L., P.Y., W.J. and I.Z. discuss the results. All authors discussed the results and commented on the manuscript. T.N., D.C.R. D.Y. and T.Z. directed the research.

**Competing interests:** All authors declare no competing interests.

**Data availability:** All data needed to evaluate the conclusions in the paper are present in the paper and/or the Supplementary Information. The data that support the findings of this study are also available in Zenodo with the identifier 10.5281/zenodo.21284586.

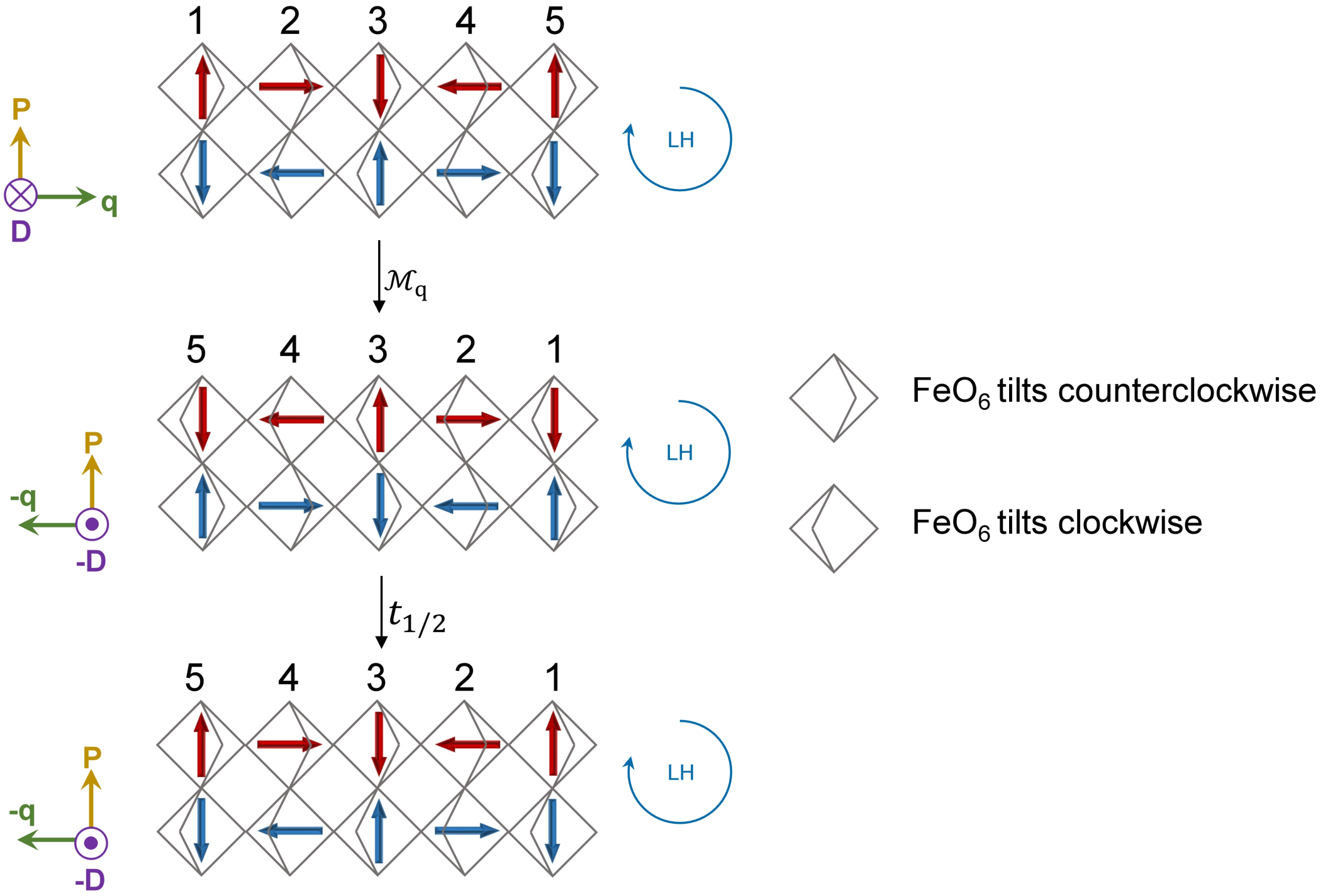

**Extended Data Figure 1. Conserved operation** $\boldsymbol{t_{1/2}\mathcal{M}_q}$. The red and blue arrows represent the antiparallel coupled magnetic sublattices (the canted magnetization is ignored to simplify the schematic). The numbers label the row of $FeO_6$ octahedra. The left-handedness (LH) chirality of the cycloidal spin texture is indicated by the blue arrows.

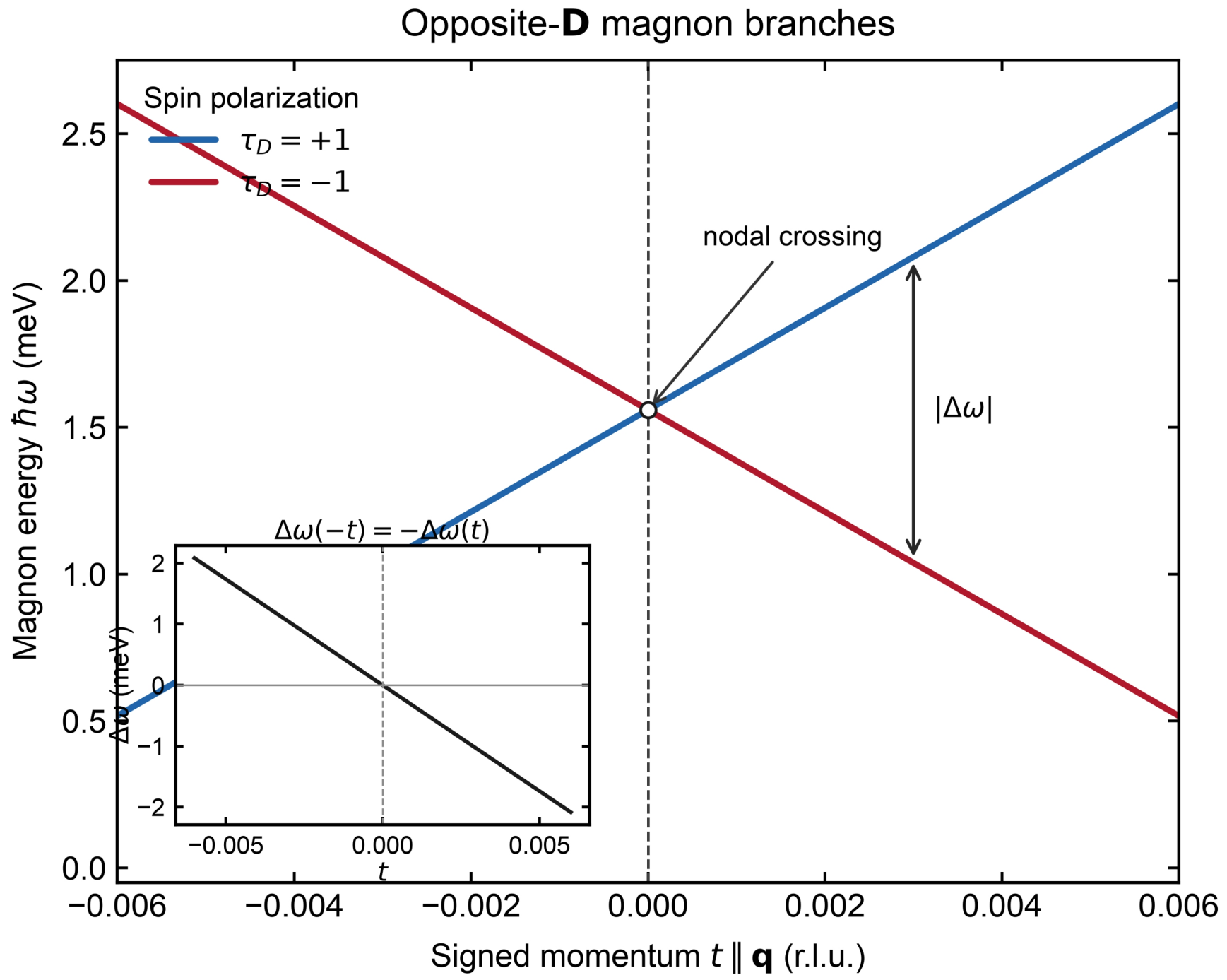

**Extended Data Figure 2. Calculated low-energy magnon branches with opposite spin polarization $\boldsymbol{\tau_D} = \pm \mathbf{1}$ along $\widehat{\boldsymbol{D}}$**. Their degeneracy at $t = 0$ is the symmetry-protected nodal crossing; away from the node, the vertical separation is $|\Delta\omega|$. The inset shows directly that $\Delta\omega(-t) = -\Delta\omega(t)$. Since $D_z \neq 0$ in both device domains, the two branches also have finite and opposite $z$-spin projections (as given by the projection relation above).

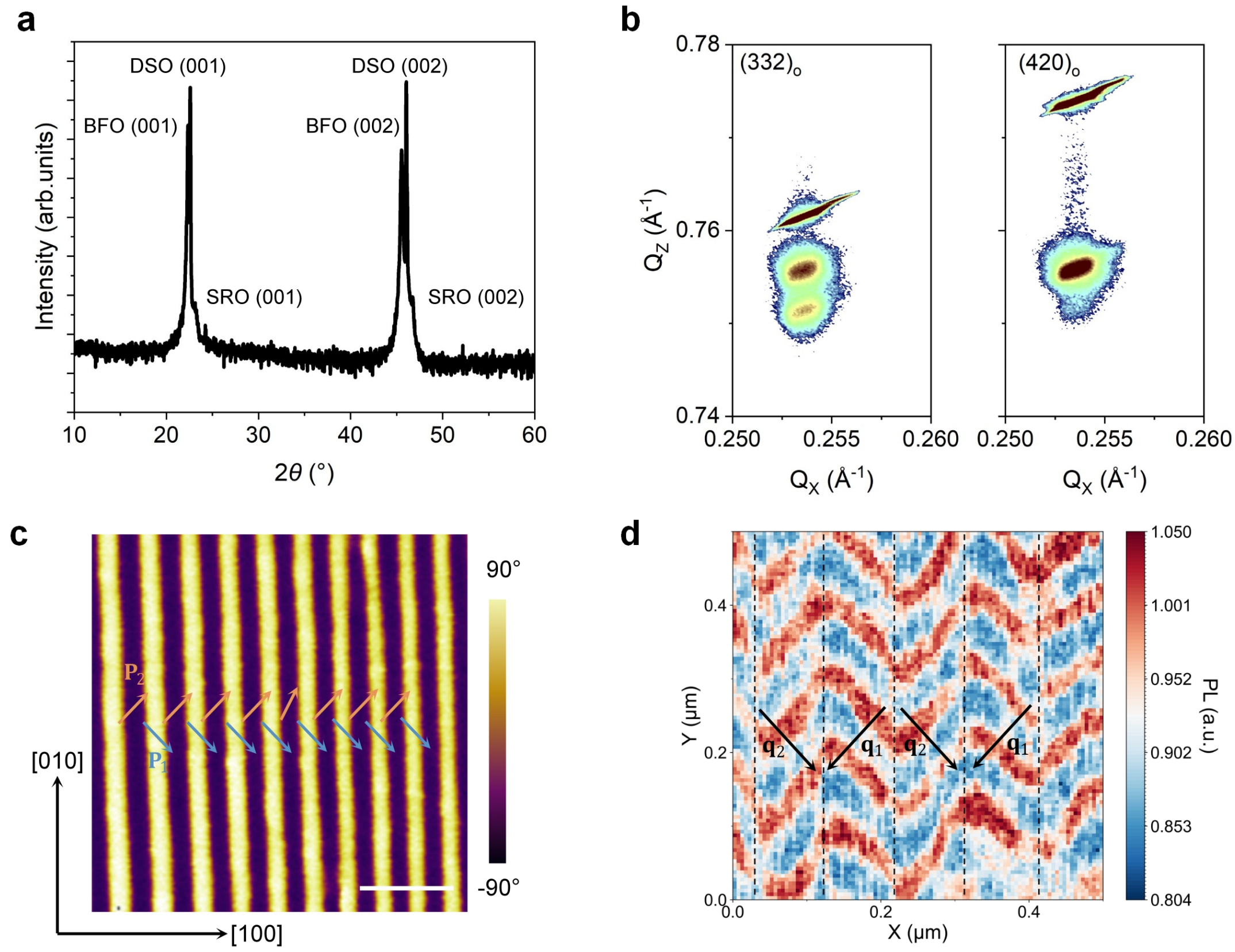

**Extended Data Figure 3. Crystalline and magnetic structure characterization. a**, X-ray diffraction pattern of fabricated BFO/SRO heterostructure on a DSO substrate. The pseudo-cubic diffraction peaks are labelled. **b**, RSM images of orthorhombic (o) $(332)_o$ and $(420)_o$ diffraction peaks, respectively. **c**, PFM image of the two-variant ferroelectric domain structure in BFO. The in-plane polarization direction in each domain is labelled. Scale bar, 500 nm. **d**, Nitrogen-vacancy magnetometry image of the cycloidal spin texture in the same two-variant BFO domain structure. Dashed lines indicate ferroelectric domain walls, and the cycloidal propagation direction in each domain is labelled.

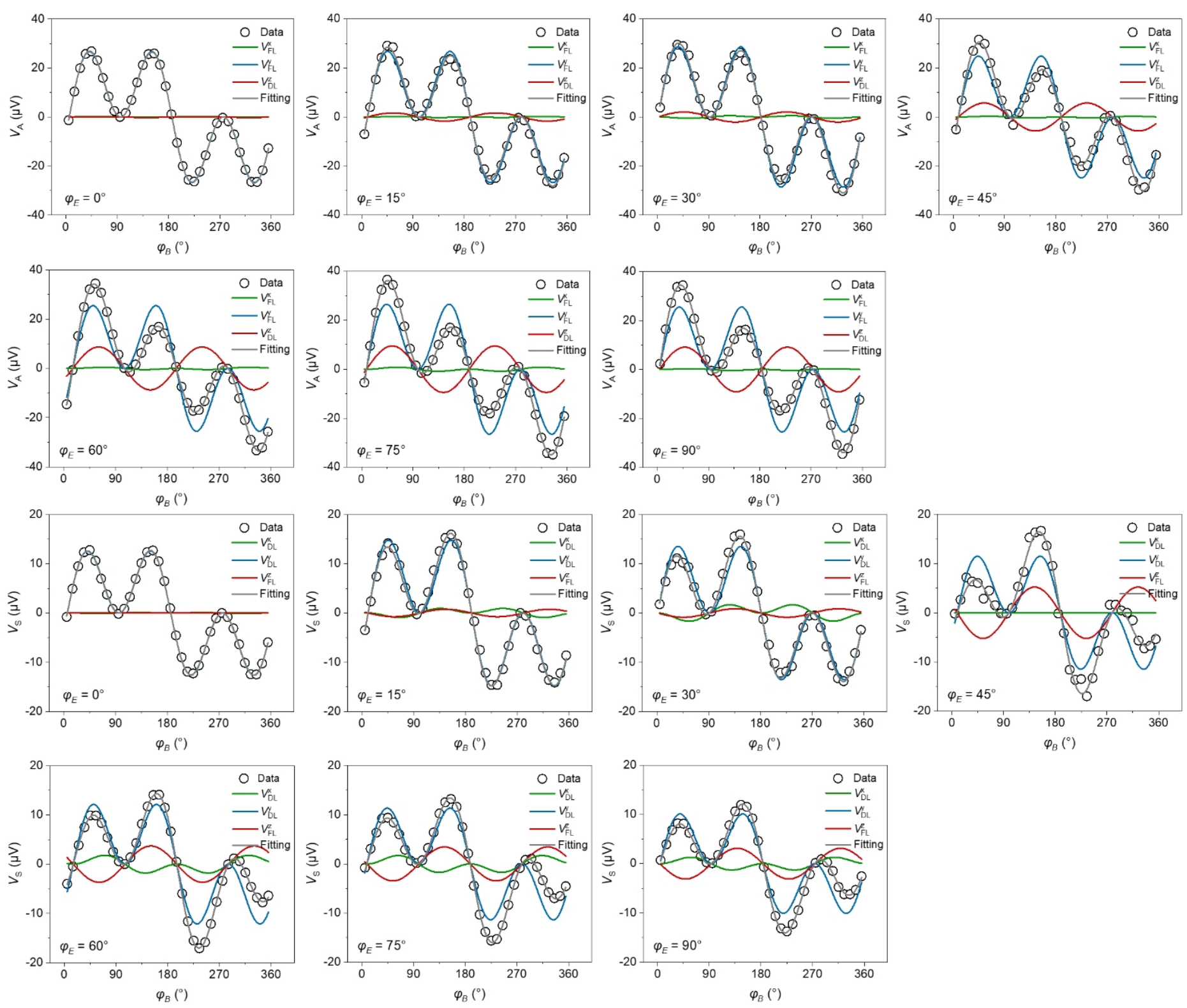

**Extended Data Figure 4**. **Representative magnetic field angular dependence of antisymmetric and symmetric components for all ST-FMR devices made from 11 nm SRO/80 nm BFO/5 nm Py heterostructures, at 6 GHz.** Both components are fitted as a function of $\varphi_B$ according to Eqs. (30-31), to extract $V_{DL}^{y}$ and $V_{DL}^{z}$ and to estimate the conventional and unconventional torque efficiencies according to Eqs. (S1-S2).

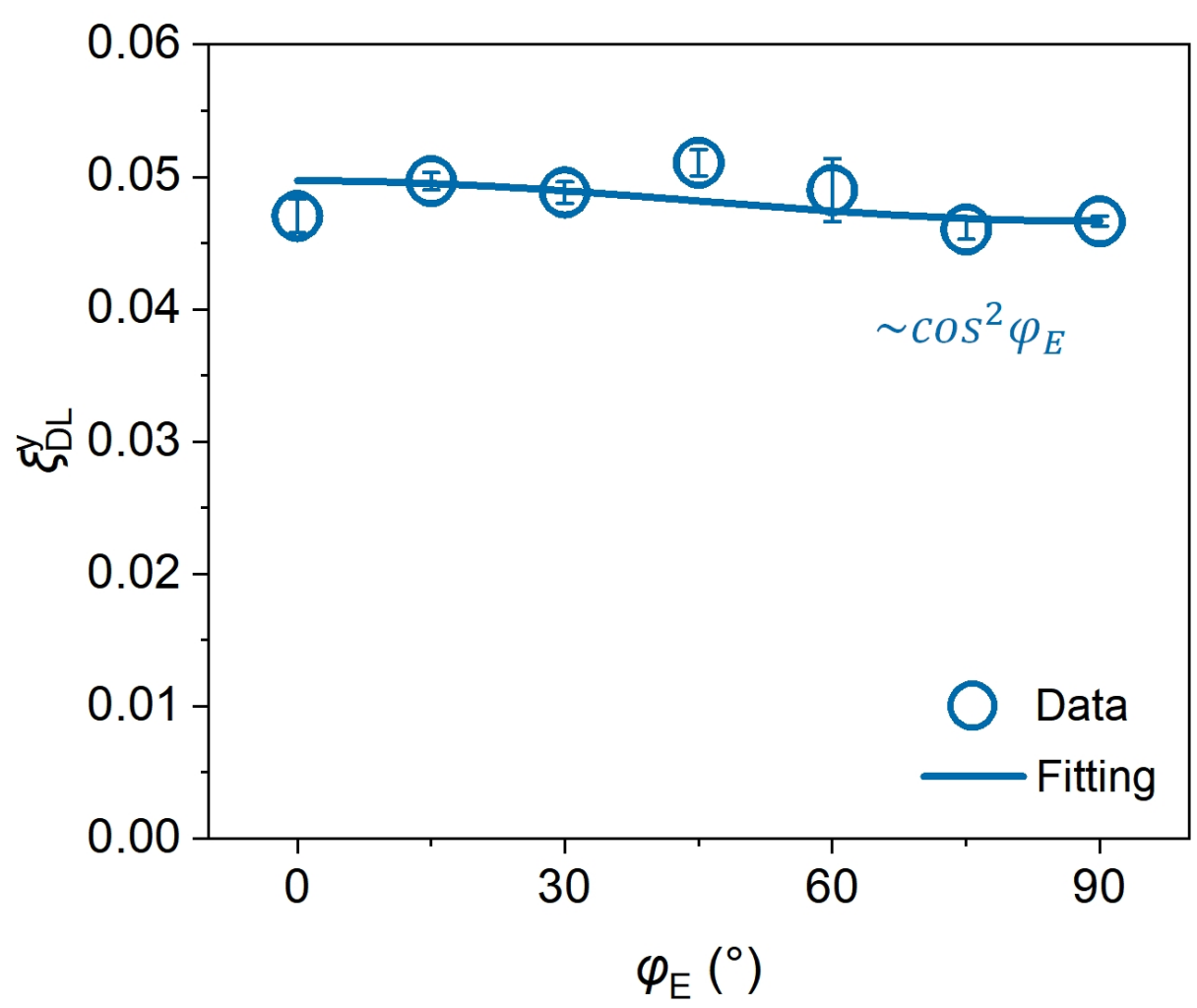

**Extended Data Figure 5. Extracted $\xi^{y}_{DL}$ as a function of $\varphi_E$ in 11 nm SRO/80 nm BFO/5 nm Py heterostructures.** $\xi^{y}_{DL}$ is fitted to a $cos^2\varphi_E$ dependence. Error bars represent the standard error of torque efficiencies extracted for different frequencies.

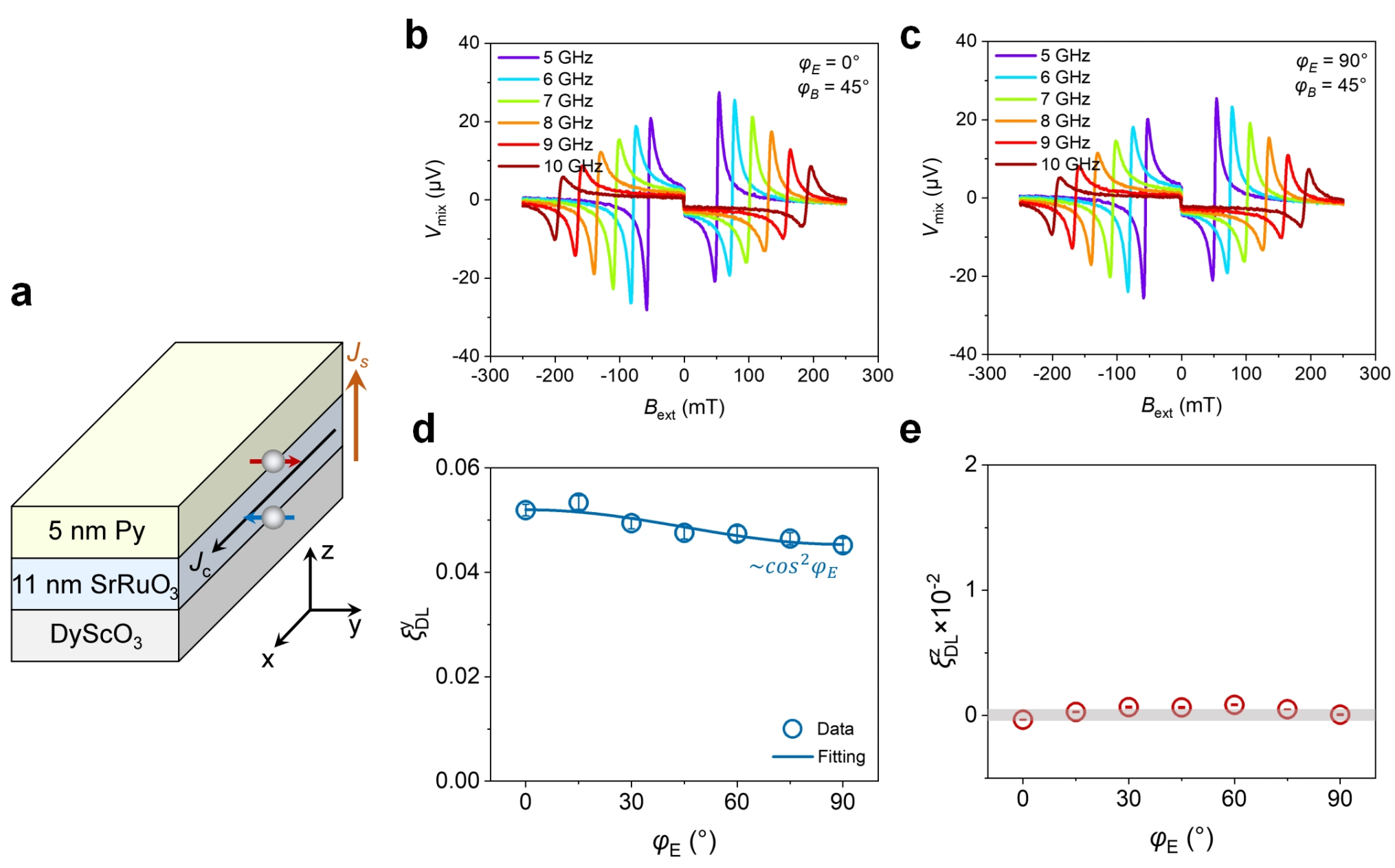

**Extended Data Figure 6. Absence of out-of-plane unconventional spin polarization generated by a $SrRuO_3$ layer. a**, To exclude any out-of-plane unconventional spin polarization generated by the $SrRuO_3$ (SRO) spin-source layer, 11 nm SRO/5 nm Py heterostructures were deposited on a $DyScO_3$ substrate. A charge current $J_c$ applied along $x$-direction generates an electron-mediated spin current $J_s$ propagating along z, carrying spin polarization $\boldsymbol{\sigma}_s$ parallel to $y$-direction. The gray spheres represent electrons. **b** and **c**, ST-FMR spectra for devices with $\varphi_E = 0°$ (**b**) and $\varphi_E = 90°$ (**c**), consistent with a conventional torque $\boldsymbol{\tau}_{DL}^{y}$, where the symmetric $V_S$ and antisymmetric $V_A$ voltage Lorentzian components both simply reverse polarity upon magnetic-field reversal. **d**, Extracted conventional torque efficiency $\xi_{DL}^{y}$ follows a weak $cos^2\varphi_E$ modulation associated with the residual crystalline anisotropy imposed by orthorhombic substrate. **e**, The extracted unconventional torque efficiency $\xi_{DL}^{z}$ is negligible, within the noise level. The out-of-plane unconventional spin polarization is forbidden in SRO layer at room temperature.

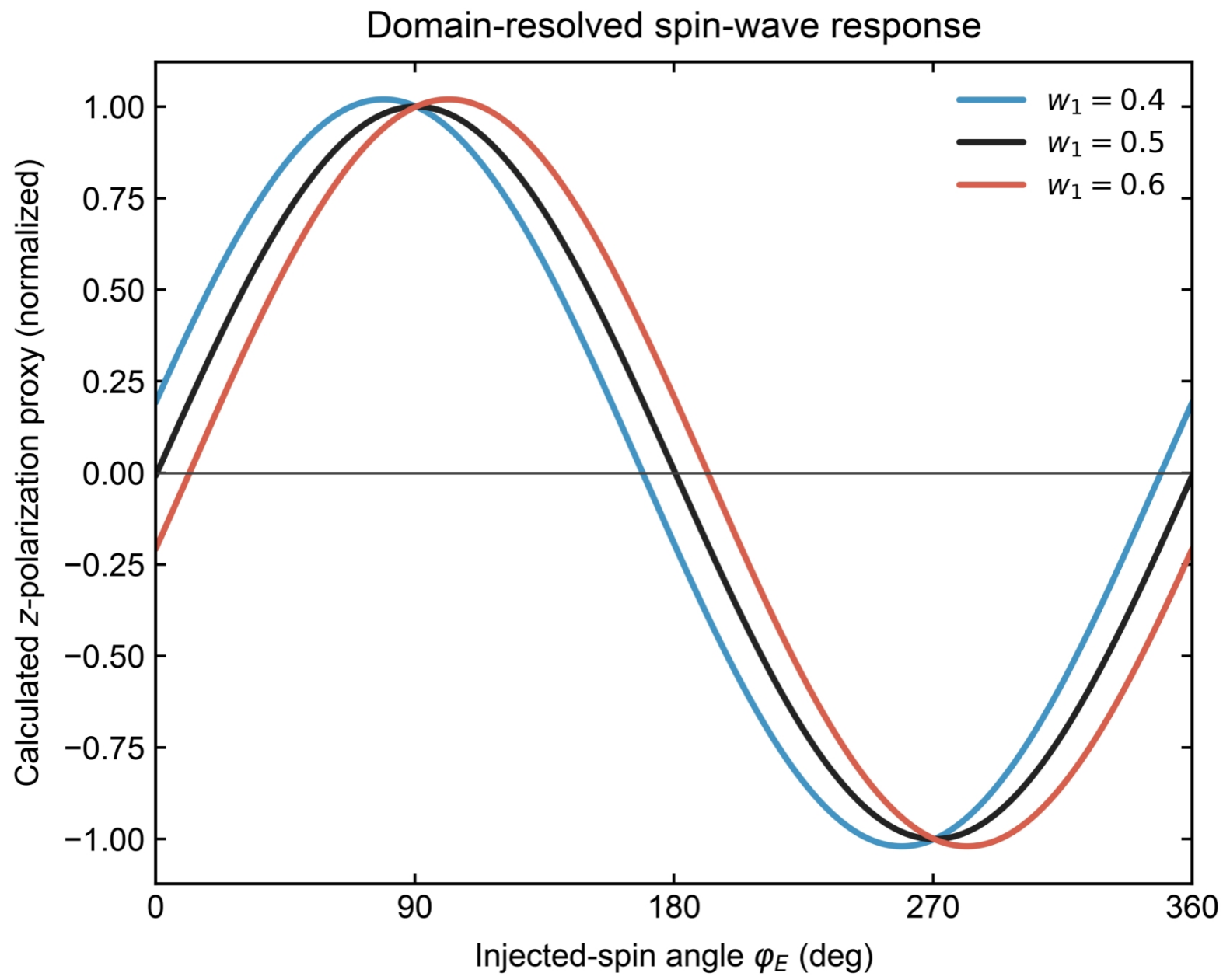

**Extended Data Figure 7**. **Normalized $z$-polarization response calculated from the domain-resolved low-energy chiral spectral vectors for various domain weights**. The calculation results are displayed with $w_1 = 0.4$, 0.5 and 0.6 for domain 1, where $w_1$ is the domain-1 weight. Equal domains give a nearly $sin\varphi_E$ dependence, with a small residual caused by the distorted lattice metric.

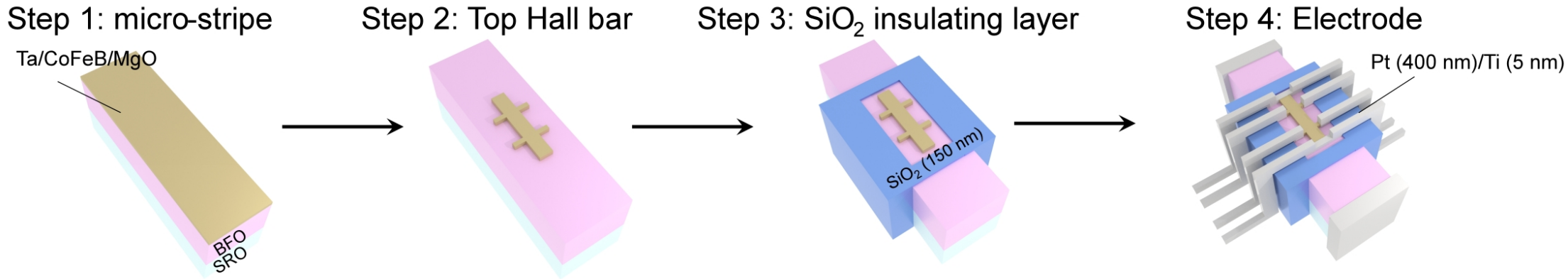

**Extended Data Figure 8. Schematic of *in-situ* device fabrication for measurements of voltage-controlled magnon spin splitting.** Step 1, The SRO/BFO/Ta/CoFeB/MgO heterostructures were patterned into 12 µm-wide and 34 µm-long microstrips. Step 2, the top Ta/CoFeB/MgO magnetic multilayer is defined into an 1 µm-wide and 20 µm-long Hall-bar geometry for anomalous Hall detection. Step 3, a 150-nm-thick $SiO_2$ insulating layer is deposited and patterned to electrically isolate the top Hall circuit from the bottom SRO channel. Step 4, 5-nm Ti/400-nm Pt electrodes are deposited to provide separate electrical contacts.

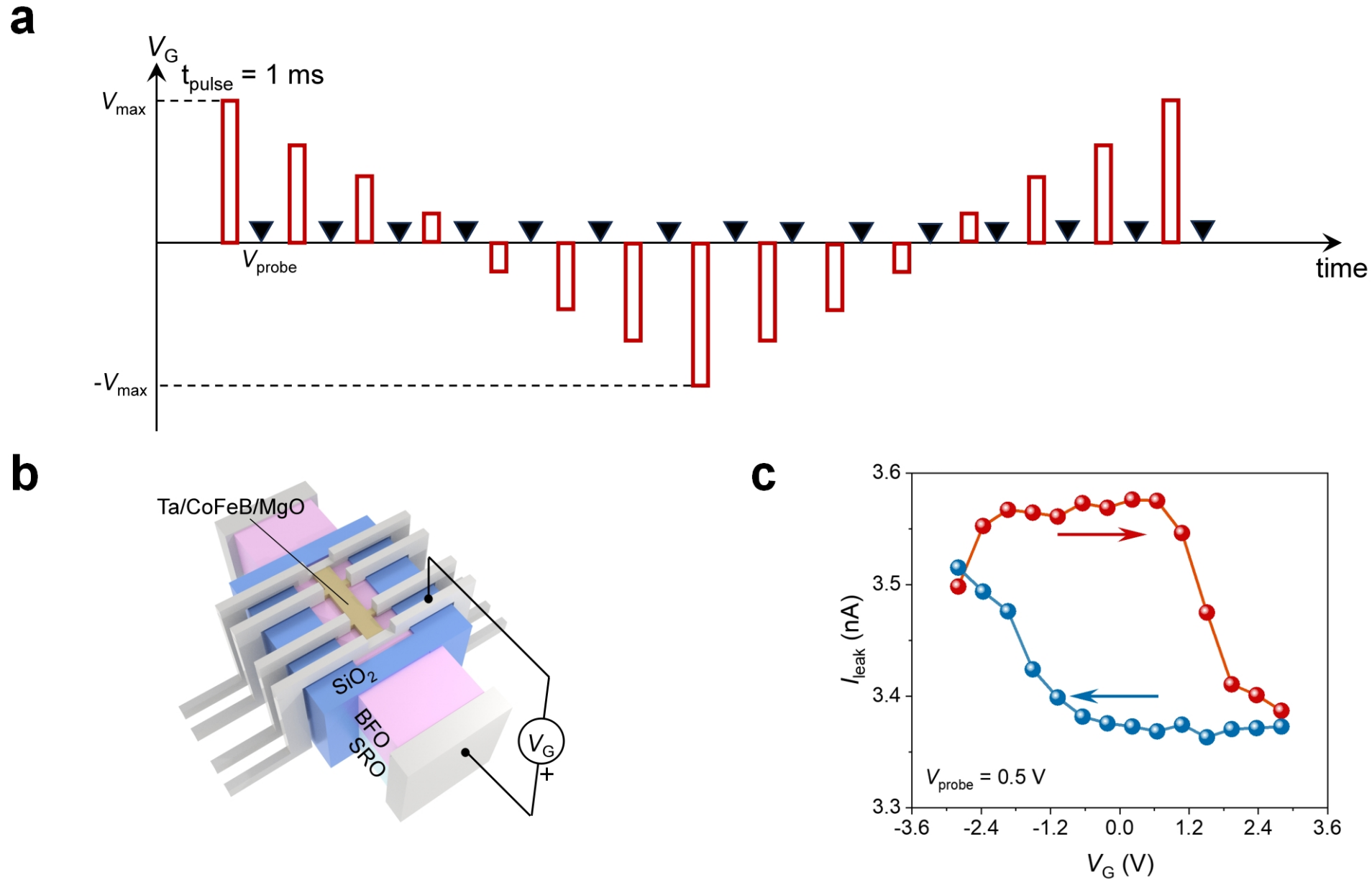

**Extended Data Figure 9. Verification of ferroelectric polarization switching via the ferroelectric diode effect. A**, Schematic of the measurement timing sequence. A series of gate voltage pulses ($V_G$) are applied with a pulse width of 1 ms. Following the removal of $V_G$ and a 4 s stabilization delay, a small readout voltage $V_{\text{probe}}$ is applied to monitor the remanent leakage current ($I_{\text{leak}}$). **B**, Schematic of the experimental setup, where the positive voltage is defined from the bottom to the top electrode. **C**, Recorded $I_{\text{leak}}$ hysteresis loop obtained by sweeping $V_G$ between $+2.8$ V to $-2.8$ V. A constant probe voltage of $V_{\text{probe}}$ = 0.5 V is applied to detect the leakage current between gate voltage pulses.

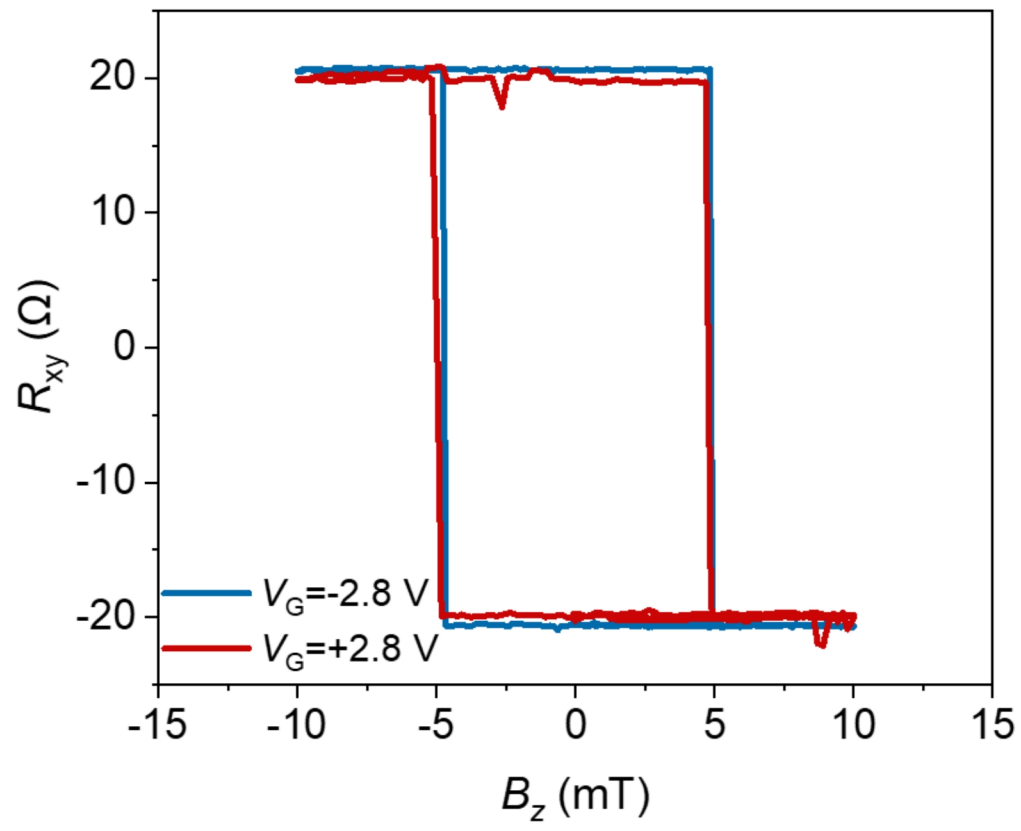

**Extended Data Figure 10. Absence of voltage-controlled magnetic anisotropy (VCMA) effect.** The anomalous Hall resistance loops were measured after applying a gate voltage $V_G = \pm 2.8$ V to set the two opposite ferroelectric states of BFO. The nearly identical loop shapes and coercive fields indicate that ferroelectric switching does not measurably alter the perpendicular magnetic anisotropy of CoFeB, excluding VCMA as the origin of the gate-dependent reversal of the Hall-loop shift.

| Physical quantities / symmetries | | P | q | D | handedness | k ∥ P | k ∥ q | k ∥ D | σ ∥ P | σ ∥ q | σ ∥ D |
|---|---|---|---|---|---|---|---|---|---|---|---|
| operations | $t_{1/2}\mathcal{M}_q$ | + | − | − | + | + | − | + | − | + | − |
| | $\mathcal{P}$ | − | − | + | − | − | − | − | + | + | + |
| | $\mathcal{T}$ | + | + | + | + | − | − | − | − | − | − |

**Extended Data Table 1. The symmetry operations on physical quantities including P, q, D, handedness as well as reciprocal-space physical quantities magnon momentum k and magnon spin polarization σ components parallel to P, q, D.** The entries in the table represent whether the quantity is even (+) or odd (−) under the symmetry operations.

| Magnetic order | Representative Materials | Magnetic structure | Parity | Excitation | Evidence | Voltage programming | Temperature | References |
|---|---|---|---|---|---|---|---|---|
| Collinear compensated antiferromagnet | $RuO_2$, CrSb, FeS, MnTe | | Even | Electron<br>Magnon | Electrical transport<br>Neutron scattering<br>ARPES spectrum | Not demonstrated | RT | 4-6, 22, 64-66 |
| Noncollinear Compensated antiferromagnet | $Mn_3X$ (X = Sn, Pt, Ir, etc.) | | Not defined | Electron | Electrical transport<br>Optical response | Not demonstrated | RT | 67-69 |
| | $MnTe_2$ | | Even | Electron | ARPES spectrum | Not demonstrated | Cryogenic temperature | 15 |
| | $Gd_3(Ru_{1-\delta}Rh_\delta)_4Al_{12}$ | | Odd | Electron | Electrical transport | Not demonstrated | Cryogenic temperature | 13 |
| | $NiI_2$ | | Odd | Electron | Photocurrent | Demonstrated | Cryogenic temperature | 14 |
| | $BiFeO_3$ | | Odd | Magnon | Magnon transport | Demonstrated | RT | This work |

: spin sublattices : Different crystal environment : Different non-magnetic atoms layer : spiral plane

**Extended Data Table 2. Comparison of spin splitting phenomena in compensated antiferromagnets.** Selected collinear and noncollinear compensated antiferromagnets are compared according to their magnetic structure, momentum parity of spin splitting, excitation, evidence, voltage programming and operating temperature. This work combines handedness-driven odd-parity magnon spin splitting, magnon-transport detection, nonvolatile voltage control and room-temperature operation. Arrows denote magnetic sublattices; diamonds denote distinct crystalline environments; triangles denote different nonmagnetic atomic layers; and ellipses indicate spiral planes. ARPES, angle-resolved photoemission spectroscopy; RT, room temperature; N/A, not applicable.

Supplementary Materials for

# Ferroelectric switching of odd-parity magnon spin splitting

Yuhan Liang[1,2], Xingyu Yan[3], Bowen Hao[4,6], Ziye Zhu[4,6], Tianle Sui[3], Daniel Pharis[2], Xiaoxi Huang[2], Rakshit Jain[2,8], Tong Zhou[4,6†], Di Yi[3,†], Wanjun Jiang[5], Pu Yu[5], Igor Žutić[7], Yuan-Hua Lin[3], Daniel C. Ralph[2,8,†], Tianxiang Nan[1,†]

1 *School of Integrated Circuits and Beijing National Research Center for Information Science and Technology (BNRist), Tsinghua University, Beijing 100084, China*

2 *Department of Physics, Cornell University, Ithaca, NY 14853, USA*

3 *School of Materials Science and Engineering, Tsinghua University, Beijing 100084, China*

4 *Ningbo Institute of Digital Twin, Eastern Institute of Technology, Ningbo, Zhejiang 315200, China*

5 *Department of Physics, Tsinghua University, Beijing 100084, China*

6 *Department of Physics, University of Science and Technology of China, Hefei, Anhui 230026, China*

7 *Department of Physics, University at Buffalo, State University of New York, Buffalo, NY 14260, USA*

8 *Kavli Institute at Cornell for Nanoscale Science, Ithaca, NY 14853, USA*

[†] Email: tzhou@eitech.edu.cn; diyi@mail.tsinghua.edu.cn; dcr14@cornell.edu; nantianxiang@mail.tsinghua.edu.cn

**Note 1. Estimation of magnon torque efficiencies**

According to our previous studies[1], the magnon spin-torque efficiencies from the ST-FMR signals can be calculated as:

$$\xi_{DL}^{y} = \frac{V_{DL}^{y}}{\sqrt{2}} \frac{eM_s t_{Py}}{\hbar J_{rf}} \frac{4\Delta}{I_{rf} R_{AMR}} \frac{2B_0 + \mu_0 M_{\text{eff}}}{\sqrt{B_0(B_0 + \mu_0 M_{\text{eff}})}} \quad \text{(S1)}$$

$$\xi_{DL}^{z} = \frac{V_{DL}^{z}}{\sqrt{2}} \frac{eM_s t_{Py}}{\hbar J_{rf}} \frac{4\Delta}{I_{rf} R_{AMR}} \frac{2B_0 + \mu_0 M_{\text{eff}}}{B_0 + \mu_0 M_{\text{eff}}} \quad \text{(S2)}$$

where $M_s$ is the saturated magnetization and $M_{\text{eff}}$ is the effective magnetization. We take $M_s \approx M_{\text{eff}}$ due to the in-plane anisotropy of the Py layer. $t_{Py}$ = 5 nm is the thickness of the Py layer. $J_{rf}$ is the rf current density flowing through the SRO layer, $I_{rf}$ is the rf current flowing through whole device, and $R_{AMR}$ is the anisotropic magnetoresistance (AMR). $\Delta$ is the line width of the ST-FMR spectrum, and $B_0$ is the resonance field. Here a factor of $\sqrt{2}$ is introduced because the ST-FMR spectrum is measured using a lock-in amplifier with 100% modulation of the microwave power[2].

The raw data frequency dependence of the resonant field $B_0$ as determined by ST-FMR measurements using frequencies from 5 to 10 GHz for the devices with $\varphi_E = 0°$ and 90° are shown in Fig. S1a and Fig. S1b, respectively. The effective demagnetization $\mu_0 M_{\text{eff}}$ of the Py layer can be determined by a fit to the Kittel formula[3], expressed as $f = (\gamma/2\pi)\sqrt{B_0(B_0 + \mu_0 M_{\text{eff}})}$, where $\gamma$ is the gyromagnetic ratio (Fig. S1c). The values of $\mu_0 M_{\text{eff}}$ for twelve different devices with applied current angles $\varphi_E$ from 0° to 90° are shown in Fig. S1d, and the estimated $\mu_0 M_{\text{eff}}$ = 0.48 T.

To calibrate $I_{rf}$ passing through the devices for a given ST-FMR measurement power and to estimate $J_{rf}$ within the SRO layer we compare the degree of sample heating due to $I_{rf}$ with the heating due to a direct current $I_{dc}$. Here, we elucidate the process by taking the ST-FMR device with $\varphi_E = 90°$. First, we monitored $R_{xx}$ by applying a small probe dc current $I_{probe}$ = 0.1 mA while gradually increasing the $I_{RF}$ by increasing microwave power from 0 to 15 dBm at frequency of 6 and 7 GHz, as shown in Fig. S2a. $R_{xx}$ increases with the enhancement of microwave power, which can be ascribed to a Joule heating effect. Then we removed $I_{RF}$ and monitored $R_{xx}$ by gradually increasing $I_{dc}$ up to ±2 mA, as shown in Fig. S2b. The monitored $R_{xx}$ can be well fitted by $R_{xx} = R_0 + R_1 I_{dc}^2$, following a Joule heating model. By comparing $R_{xx}$ for microwave power of 15 dBm with the fitting curve and assuming $R_{xx} = R_0 + R_1(\frac{I_{rf}}{\sqrt{2}})^2$, we calibrated $I_{rf}$ to be 2.26 mA for 6 GHz and 6.27 mA for 7 GHz, respectively, see Fig. S2c. Here we ignore the thermal effect induced by $I_{probe}$. We also characterized the AMR of the ST-FMR device with $\varphi_E = 0°$ by applying an external magnetic field $\mathbf{B}_{ext}$ = 0.1 T at various angles $\varphi_B$. A small dc current $I_{dc}$ = 0.1 mA was applied to monitor $R_{xx}$. The results are shown in Fig. S2d. $R_{xx}$ can be well fitted by $R_{xx} = R_0 + R_{AMR} cos^2\varphi_B$. Here we found $R_{AMR}$ = 2.85 Ω for the ST-FMR device with $\varphi_E = 90°$ . The values of $R_{AMR}$ for all devices are shown in Fig. S2e.

We then calculated $J_{rf}$ flowing through the $SrRuO_3$ layer by using a parallel resistance model. An 11 nm $SrRuO_3$ thin film on a $(110)_o$-oriented DSO substrate and a 5 nm Py/3 nm Ti heterostructure on an insulating $SiO_2$/Si substrate were both fabricated into Hall bar device for resistivity measurements. The resistivity of both samples was monitored by applying 0.1 mA dc current as a function of temperature from 300 K to 10 K, and the results are shown in Fig. S2f. The resistivity of the 11 nm SRO ($\rho_{SRO}$) at 300 K is about 421 μΩ·cm. A kink appears around 150 K due to the ferromagnetic-paramagnetic phase transition of $SrRuO_3$. Assuming the top 3 nm Ti oxidized in air, the resistivity of the 5 nm Py ($\rho_{Py}$) is about 70 μΩ·cm. Thus, we can calculate $J_{rf}$ as:

$$J_{rf} = \frac{I_{RF}\rho_{Py}}{\rho_{SRO}t_{Py} + \rho_{Py}t_{SRO}}\frac{1}{w} \tag{S3}$$

where $w$ = 16 μm is the width of the ST-FMR devices. Here we calculate $J_{rf}$ = $1.5\times10^{10}$ A/m$^2$ for ST-FMR device with $\varphi_E = 90°$. Having established this method, we calibrated $J_{rf}$ and $R_{AMR}$ for all the devices with various $\varphi_E$, and then calculated $\xi_{DL}^{y}$ and $\xi_{DL}^{z}$ after collecting all required parameters for Eqs. (S1-S2). The extracted unconventional torque efficiency $\xi_{DL}^{z}$ follows the predicted $sin\varphi_E$ dependence (Fig. 2g in main text), whereas $\xi_{DL}^{y}$ is nearly isotropic, with only a weak $cos^2\varphi_E$ modulation associated with the residual crystalline anisotropy of the SRO spin source (Extended Data Figure 5).

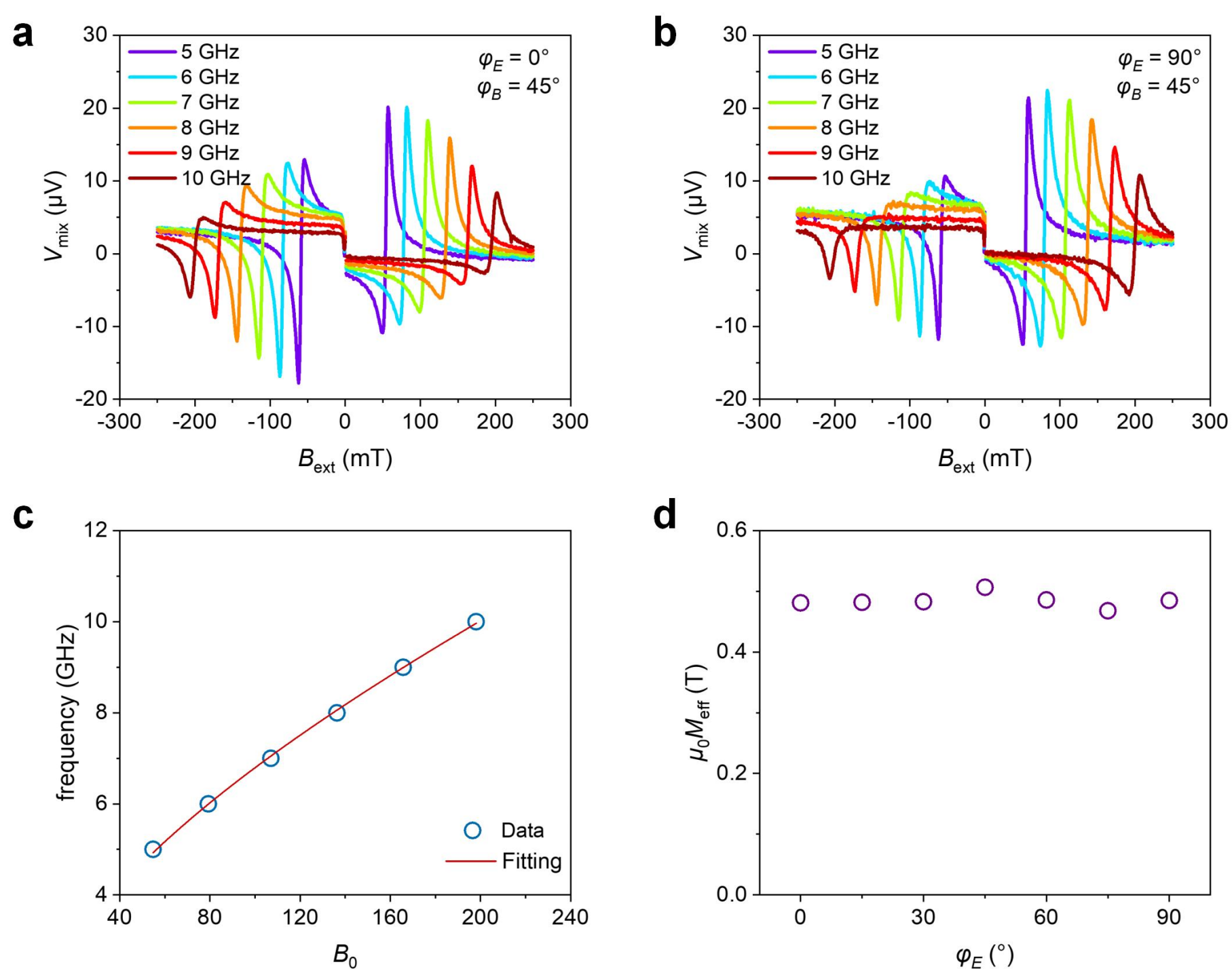

**Supplementary Figure 1. Details of ST-FMR spectra for 11 nm SRO/80 nm BFO/5 nm Py heterostructures. a** and **b,** ST-FMR spectra for devices with $\varphi_E = 0°$ (**a**) and $\varphi_E = 90°$ (**b**) from 5 GHz to 10 GHz. **c**, Frequency as a function of resonant field, with a fit to the Kittel formula (red solid line). **d**, Effective magnetization $\mu_0 M_{\mathrm{eff}}$ determined for twelve devices with different angles of applied current $\varphi_E$.

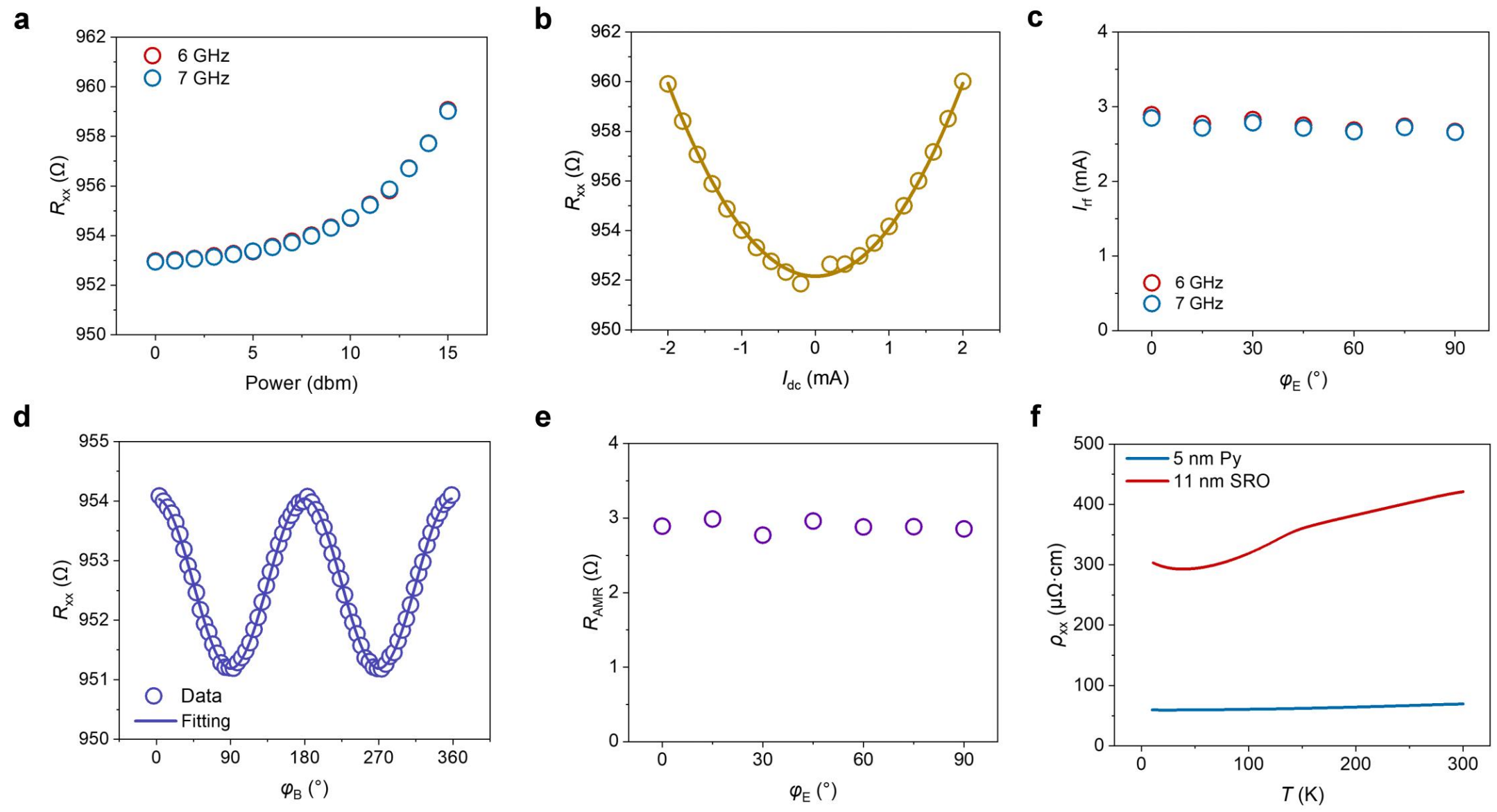

**Supplementary Figure 2. Calibration of rf current and the AMR resistance in the ST-FMR devices made from 11 nm SRO/80 nm BFO/5 nm Py heterostructures. a**, The measured device resistance $R_{xx}$ as a function of applied microwave power with frequency of 6 and 7 GHz, respectively. A small dc current $I_{dc}$ = 0.1 mA is applied to monitor the resistance. **b**, The measured device resistance as a function of applied $I_{dc}$. The blue line represents the fit to $R_{xx} = R_0 + R_1 I_{dc}^2$. **c**, Calibrated $I_{RF}$ for STFMR devices with various $\varphi_E$. **d**, The measured AMR with a fit to $R_{xx} = R_0 + R_{AMR} cos^2 \varphi_B$ (purple line). **e**, The detected $R_{AMR}$ for STFMR devices with various $\varphi_E$. **f**, The temperature dependent resistivity of 11 nm SRO and 5 nm Py, respectively.

## Supplementary References